\documentclass{LMCS}
\usepackage{graphicx,amssymb, amsmath, amssymb,latexsym, color,ifpdf}
\usepackage{enumerate,hyperref}

\newtheorem{notat}[thm]{Notation}
\def\eqalign#1{\null\,\vcenter{\openup\jot\mathsurround=0 pt
  \ialign{\strut\hfil$\displaystyle{##}$&$\displaystyle{{}##}$\hfil
      \crcr#1\crcr}}\,}

\newcommand{\ar}[1]{\stackrel{{#1}}{{\longrightarrow}}}
\newcommand{\Ar}[1]{\stackrel{{#1}}{\Longrightarrow}}
\newcommand{\alt}{\raisebox{-0.22cm}{ \rule{0.2mm}{0.6cm} } }

\newcommand{\rid}{\ar {\tau}}

\newcommand{\remove}[1]{{}}

\newcommand{\bigfrac}[2]{
\renewcommand{\arraystretch}{1.6}
\begin{array}{c}#1\\
\hline #2
\end{array}}

\newcommand{\may}{\:\mbox{{\it may}}\:}
\newcommand{\must}{\:\mbox{{\it must}}\:}
\newcommand{\fair}{\:\mbox{{\it fair}}\:}

\newcommand{\sfmust}{\:\mbox{{\it sfmust}}\:}
\newcommand{\wfmust}{\:\mbox{{\it wfmust}}\:}
\newcommand{\notmust}{\:\:\not\!\!\!\!  \!\!\must}
\newcommand{\notfair}{\:\:\not\!\!\!\!  \!\!\fair}
\newcommand{\notsfmust}{\:\:\not\!\!\!\!  \!\!\sfmust}
\newcommand{\notwfmust}{\:\:\:\not\!\!\!\!\!\!\!\!\wfmust}

\newcommand{\topp}{\mbox{\it top}}
\def\N{\mathbb{N}}

\def\doi{5 (2:15) 2009}
\lmcsheading%
{\doi}
{1--27}
{}
{}
{Jul.~10, 2008}
{Jun.~22, 2009}
{}   

\begin{document}

\title[Explicit fairness in testing semantics]{Explicit fairness in testing semantics}

\author[D.~Cacciagrano]{Diletta Cacciagrano\rsuper a}	
\address{{\lsuper a}Dipartimento di Matematica e Informatica, 
Universit\`a degli Studi di Camerino, Camerino, Italy}	
\email{\{diletta.cacciagrano,flavio.corradini\}@unicam.it}
\thanks{{\lsuper{a,b}}The work of Diletta Cacciagrano and Flavio
    Corradini has been supported by the Investment Funds for Basic
    Research (MIUR-FIRB) project Laboratory of Interdisciplinary
    Technologies in Bioinformatics (LITBIO) and by Halley
    Informatica.} 

\author[F.~Corradini]{Flavio Corradini\rsuper b}	
\address{\vskip-6 pt}	

\author[C.~Palamidessi]{Catuscia Palamidessi\rsuper c}
\address{{\lsuper c}INRIA Futurs and LIX, \'Ecole Polytechnique, France}	
\email{catuscia@lix.polytechnique.fr}
\thanks{{\lsuper c}The work of Catuscia Palamidessi has been partially supported by the INRIA DREI 
\'Equipe Associ\'ee PRINTEMPS and by the INRIA ARC project ProNoBiS}


\begin{abstract}
In this paper we investigate fair computations in the $\pi$-calculus
\cite{MPW92}. Following Costa and Stirling's approach for CCS-like
languages \cite{CS84,CS87}, we consider a method to label process actions 
in order to filter out unfair computations. We contrast the existing
fair-testing notion \cite{RV07,NC95} with those that naturally arise by
imposing weak and strong fairness.
This comparison provides insight about the expressiveness of the various `fair' testing 
semantics and about their discriminating power.
\end{abstract}

\maketitle

\section{Introduction}

One of the typical problems of concurrency is to  ensure that all the tasks that are supposed to be executed do not get postponed indefinitely in favor of other activities. This property, which is called \emph{fairness}, can be implemented by using a particular scheduling policy that excludes unfair behavior. For instance, in Pict \cite{PT00}, (weak) fairness is obtained by using FIFO channel queues and a round-robin policy for process scheduling. A stronger property (strong fairness) is obtained by using priority queues. 

Of course in practice it is not feasible to impose that all implementations adopt a certain scheduler. One reason is  that, depending on the underlying machine, one scheduling policy  may be much more efficient than another one. Hence fairness has been studied, since the beginning of  the research on Concurrency,  as an abstract property and independently from the implementation. \vfill\eject

\subsection{Fairness in literature}
Most of the common notions of fairness 
share the same general form: ``Every {\it entity} that is enabled {\it sufficiently often} will eventually make progress.'' 
Varying the interpretations of `{\it entity}' and `{\it sufficiently often}' leads to different notions of fairness. 

Kuiper and de Roever \cite{KdR83} identified a wide hierarchy of  
fairness notions for the CSP language (channel fairness, process fairness, guard fairness,
and communication fairness), according to the entity 
taken into account (respectively channel, process, guard and communication). Each of these fairness notions 
have a weak and a strong variant,
which differ in the interpretation of {\it sufficiently often}: weak forms of fairness are concerned
with continuously enabled entities, whereas strong forms of fairness are concerned with the
infinitely enabled entities. 

Independently, Costa and Stirling investigated (weak and strong) fairness of actions
for a CCS-like language without restriction in \cite{CS84}, and fairness of components 
for the full CCS in \cite{CS87}.  An important result of their investigation was 
the characterization of fair executions in terms of the concatenation of certain finite sequences, called LP-steps. 
This result allowed expressing fairness as a local property 
instead than  a property of complete maximal executions. 

Although  \cite{KdR83} and \cite{CS84, CS87}  seem to define different  fairness  varieties, 
there is a correspondence between some notions in the two approaches   (up to the  language 
on which the study is based): 
guard fairness corresponds to fairness of actions, while process 
fairness corresponds to fairness of components. 
However, the communication mechanism  of the languages chosen for the study - CSP in \cite{KdR83} and CCS in \cite{CS84, CS87} - modifies 
the interrelationships among notions. In fact, in CSP processes communicate by name, each channel corresponds precisely to a
pair of processes, i.e only two processes communicate along any given channel and only one
channel is used between any two processes;  on the other hand, in CCS  any number of processes may communicate along a 
given channel, and two processes may communicate along any number of channels. This implies that some fairness notions are related in CSP while they
are not related in CCS. 
For example, while every channel-fair computation 
is also process-fair in CSP (\cite{Fra86}), in CCS it is possible for a particular channel 
to be used sufficiently  often and yet for another process to become blocked while trying to use 
that same channel\footnote{It suffices to 
consider the term $\bar a\: |\: !a.\bar a\: |\: \bar a $, where $a$ and  $\bar a$ denote actions of input and output on channel $a$, respectively, and  $!a.\bar a$ denotes 
a process which can perform infinitely often 
an input on channel $a$, followed by an output on the same channel. 
Although channel $a$ must be used
infinitely often along any infinite computation, it is possible under 
{\it channel fairness} that the leftmost $\bar a$  is ignored, 
while the right-most  $\bar a$ synchronizes continually with the process $!a.\bar a$. 	
This is not the  case under process fairness.}. 

Hennessy \cite{Hen85}  introduced the concept of fairness in his 
{\em acceptance trees} model, by adding  limit points indicating which 
infinite paths are fair. The notion of fairness incorporated into this semantics is a form of
unconditional fairness: an infinite execution is considered fair if every process makes 
infinitely many transitions along that computation.

Francez  \cite{Fra86} characterized the notions of fairness in \cite{KdR83} in terms of a so-called 
{\em machine closure} property and by means of  a topological model. 

Fairness has also been investigated in the context of probabilistic systems. 
Koomen \cite{Koo85} explained fairness with probabilistic
arguments: the {\em Fair Abstraction Rule} establishes that no matter how small the
probability of success is, if one tries often enough one will eventually
succeed.  Pnueli introduced in \cite{Pnu83} the notion of {\it extreme fairness}
and $\alpha$-{\it fairness}, to abstract from the precise values of
probabilities. 

\subsection{Fairness in bisimulation equivalences and testing semantics}
Observational equivalences and preorders can have different bearings with respect to fairness. 
In particular, this is the case of testing preorders \cite{DH84} and bisimulation equivalences \cite{Mil89, Par80}.  

The first framework was presented by De Nicola and Hennessy in their  seminal work \cite{DH84},  where 
they proposed the concept of testing and defined the {\it must}- and the {\it may}-testing semantics, as well as their induced preorders. 
Given a process $P$ and a test (observer) $o$,  
\begin{enumerate}[--]
\item $P\:\may\: o$ means that there exists a successful computation from $P\: |\:o$ 
(where $|$ is the parallel operator, and successful means that there is a state where the special action $\omega$ is enabled); 
\item $P\:\must\: o$ means that every maximal computation from 
$P\: |\:o$ is successful; 
\item The preorder $P\leq_{\!\mbox{\it sat}} Q$ means 
that for any test $o$, $P\:\mbox{\it sat}\: o$ implies $Q\:\mbox{\it sat}\: o$, where  $\mbox{\it sat}$  denotes $\may$ or  $\must$; 
\item The equivalence 
$P\approx_{\!\mbox{\it sat}} Q$ means $P\leq_{\!\mbox{\it sat}} Q$ and $Q\leq_{\!\mbox{\it sat}} P$.
\end{enumerate}

The second framework \cite{Mil89, Par80} arises from the principle of (mutual) simulation of systems. 
The prime representatives of this family are bisimilarity and observation congruence \cite{Mil89}. In particular,  
weak bisimulation incorporates a particular notion of fairness: it abstracts from the
$\tau$-loops (i.e  infinite sequences of $\tau$ - or internal - actions) in which the ``normal'' behavior 
can be resumed each time after a finite sequence of $\tau$-actions.
Such a property can be useful
in practice - for instance for communication protocols in systems with lossy communication media, 
which  retransmit lost messages. There is a fairness principle implicitly associated with 
such systems, based on the assumption that the path which stays in the loop forever is not a possible behavior of the system.
Interesting proofs of protocol correctness based on this principle are given  in \cite{Bri84, LM87}. 

Bisimulation equivalences  
are usually rather strict, since they  depend on the whole branching structure of processes, 
which in some cases may be not relevant.  On the other hand, most of the standard testing 
preorders interpret
 $\tau$-loops as divergences, making them quasi-observable.  
In fact, the {\it must}-predicate on $P\: |\: o$ 
immediately fails if $P$  is able to do a $\tau$-loop  that never reaches a
successful state. Hence, while the standard testing equivalences are coarser than weak 
bisimulation in the case of divergence-free processes, they are not comparable with the
 latter in general.

In \cite{RV07} and in \cite{NC95} a new testing semantics was proposed 
to incorporate the fairness notion: the {\it fair}-testing (aka {\it should}-testing) semantics. 
In contrast to the classical {\it must}-testing (semantics), {\it fair}-testing  abstracts from certain $\tau$-loops. This is achieved by stating that  the test $o$ is satisfied if success always remains
within reach in the system under test. In other words, $P \; {\it fair }\; o$ holds if  in every 
maximal computation from $P\: |\: o$  every state can lead to success after finitely many interactions. 
The characterizing semantics for {\it fair}-testing and a similar testing scenario can already be found in \cite{Vog92}.

The relation between bisimulations and {\it fair}-testing 
was investigated  in \cite{FG98}, in the context of name-passing process calculi like the 
asynchronous $\pi$-calculus \cite{HT91}
and the join-calculus \cite{FG00}. The authors of \cite{FG98} presented a 
hierarchy of equivalences obtained as variations of Milner and Sangiorgi's weak
barbed bisimulation. In particular, they proved that the coupled barbed equivalence strictly implies the
{\it fair}-testing equivalence. They also showed that those relations coincide in
the join-calculus and on a restricted version of the asynchronous 
$\pi$-calculus, called {\it local} $\pi$-calculus, 
where reception occurs only on names bound by a restriction (not on free and received names). 

Another relation motivated by the aim of incorporating in {\it must}-testing the fairness property 
of observation congruence is the {\it acceptance}-testing, which was defined and studied in \cite{Bri88}. 
This relation is  captured by the failures model  but, in contrast to {\it must}-testing, it does not yield 
a precongruence with respect to abstraction (or hiding), a construction which
internalizes visible actions and may thereby introduce new divergences.

 The probabilistic intuitions motivating the Koomen's rule inspired another approach 
to incorporate fairness in a testing semantics \cite{NR99}.  
The authors of  \cite{NR99} defined a probabilistic {\it must}-semantics in which a 
(probabilistic) process {\it must}-satisfy a test if and only if the probability with which the process
satisfies the test equals $1$, and proved that  two non-probabilistic processes are {\it fair}-equivalent
if and only if their probabilistic versions are equivalent in the probabilistic testing semantics.

\subsection{The goal of this work: A study of testing semantics with implicit and explicit fairness}
 {\it Fair}-testing  is an appealing equivalence. Some of its advantages are that it detects deadlocks and implements fairness. It has also been used in various works. For example, \cite{BRV96} uses the {\it fair}-testing preorder  as an implementation relation for distributed communication protocols.
 
The purpose of our study is to try to make operationally explicit the fairness assumption which is 
implicit in the {\it fair}-testing semantics. The advantages of the formulation in 
operational terms is to have a better understanding of this notion. Also, it can help  eliminating some of the known drawbacks: for example,  {\it fair}-testing abstract fairness is not enforced by practical scheduling policies, and   direct proofs of equivalence are very difficult because they involve nested inductions for all quantifiers in the definition of {\it fair}-testing and all evaluation contexts.

In contrast to  \cite{NR99} we want to keep invariant the original testing scenario and try to
characterize (or approximate)  {\it fair}-testing semantics - which does not involve any probability assumption - in term of a non-probabilistic  testing semantics equipped with some explicit fairness notion.

We proceed as follows: 
\begin{enumerate}[$\bullet$]
\item We consider the choiceless $\pi$-calculus \cite{MPW92} and we develop for it 
an approach to fairness (of actions) similar to that which has been proposed in \cite{CS84, CS87} for
CCS-like languages \cite{Mil89}. More precisely, we define (i) a {\it labeling method} for $\pi$-calculus terms that  
ensures that no label occurs more than
once in a labeled term ({\it unicity}), that a label disappears only when the corresponding 
action is performed ({\it disappearance}), and that, once it has disappeared, it will not appear in the
computation anymore ({\it persistence}), (ii) the notion of {\it live action}, which refers to the fact that the action 
can currently be performed, and (iii) {\it weak} and {\it strong fairness} of actions.

\item We then contrast the existing
{\it fair}-testing semantics \cite{RV07,NC95} with those that naturally arise by
imposing weak and strong fairness \cite{CS84,CS87} on a {\it must}-testing semantics.
\end{enumerate}

\noindent In the following we justify our choices, and describe in
detail our setting and  results.

\subsection{The choiceless \texorpdfstring{$\pi$}{pi}-calculus}
The choiceless $\pi$-calculus is essentially the $\pi$-calculus without the choice operator ($+$). 
This seems a rather appealing 
framework to study fairness. In fact,  the choice operator  is a bit controversial with respect to 
fairness, because it is not clear what fairness should  mean in the case of a 
repeated execution of a choice construct. In \cite{CS87} the continuous selection of the same branch of a choice construct  turns out to be fair,  while other researcher would not agree to consider fair this kind 
of computation. The reason why it is fair in \cite{CS87} is that when the action that has not been selected comes back in the recursive call, it is considered a new action, and it is relabeled. On the contrary, in other approaches, like for instance  \cite{KdR83}, the guards that come back  are precisely the object of weak fairness.

On the other hand, thanks to the fact that the restriction operator ``$\nu$'' allows the creation of  new names and the scope extrusion, the $\pi$-calculus is more expressive than CCS, and  it is possible to represent in it various types of choices in a  compositional way by means of the parallel operator (see \cite{Nes00,NP00,Pal03}). 
In particular, the internal choice and the input-guarded choice. 
For example,  the term  $(\nu a)(\bar a \:|\:a.b.0\:| \:a.c.0)$
represents the internal choice between $b$ and $c$. If we want to repeat the execution 
 of this choice, we use the replication operator  ``!'' which creates an arbitrary number of copies of the argument.
The issue of fairness depends on where we place ``!'' in the term:   
$!(\nu a)(\bar a\:|\:a.b.0 \:| \:a.c.0)$ can produce
 an infinite sequence of ``b'''s, and the corresponding computation is considered fair because 
 the subterms  $a.b.0$, $a.c.0$ have only one copy of  $\bar{a}$ in the same scope, so if such copy synchronizes with 
 $a.b.0$, then $a.c.0$ will be disabled forever. In a sense, the term represents a new choice each time.  On the contrary,  $(\nu a) ! (\bar a\:|\:a.b.0 \:| \:a.c.0)$ 
can also  produce an infinite sequence of ``b'''s, but the corresponding computation is not fair because 
all the copies of   $\bar{a}$ are in the same scope and therefore  $a.c.0$ is always enabled. In a sense, here we repeat always the same choice.

We find that the reduction of choice to the parallel operator brings some insight to the relation between repeated choice and fairness, in the sense that the definition of fairness for the various kinds of combination of choice and repetition stems naturally from the definition of fairness for  the parallel operator. 
 
\subsection{The labeling method}
In \cite{CS84,CS87}, labels are flat sequences of $1$'s and $2$'s and are assigned 
to operators according to the syntactic structure of the term, without 
distinguishing between static and dynamic operators. 
In our approach, labels are pairs $\langle s,n\rangle$ in $(\{0,1\}^*\times \N)$ and 
are associated to prefix and replication operators; restriction and parallel operators 
do not get a label on their own. In contrast to \cite{CS84,CS87}, 
the aim is to keep separated the information about static and dynamic
operators and avoid labels which (at least for our purpose) are superfluous, thus making more 
intuitive their role in the notion of fairness.

The first component of a pair, $s$, represents the position of the
process (whose top-level operator is associated to that label) 
in the term structure, and it depends only on the (static)
parallel operator. This component ensures the
unicity of a label. The second component, $n$, provides information
about the dynamics of the process in the term structure. More precisely,
 it indicates how many actions that process has already executed since the beginning
of the computation, and it depends only on the (dynamic) prefix operator. This second component 
serves to ensure the persistence property of a label. 

Informally, a label $\langle s, n\rangle$ denotes unambiguously a parallel process - 
the one associated to $s$ - and a precise action of it - 
the one nested at level $n$ in the original term. Note that: (i) all the actions of a parallel process  
share the first label component $s$ and they only differ from the second component $n$; 
(ii) actions of different 
parallel processes at the same level  share the second 
label component $n$ and are distinguished by the first component $s$.

We give now an example to illustrate the difference with the labeling method of  \cite{CS84,CS87}.
We recall that in  \cite{CS84,CS87} the labels are assigned essentially by using the tree representing the abstract syntax of the term: we add  $1$ to the string representing the label on the left branch, and $2$ on the right branch. 

\begin{exa}\label{esa1} Consider the term 
$S\: =\:x(y).((\nu z)(z(k).0\:\: |\:\: {\bar z}h.0))\:\: |\:\: a(u).0\:$.
The left-most tree in Figure \ref{fig1} is the the labeling of $S$ in the approach of \cite{CS84, CS87}, 
while the right-most one is the the labeling of $S$ in our approach. 
\begin{figure}[t]
\begin{center}
\ifpdf
\includegraphics[width=8.75cm]{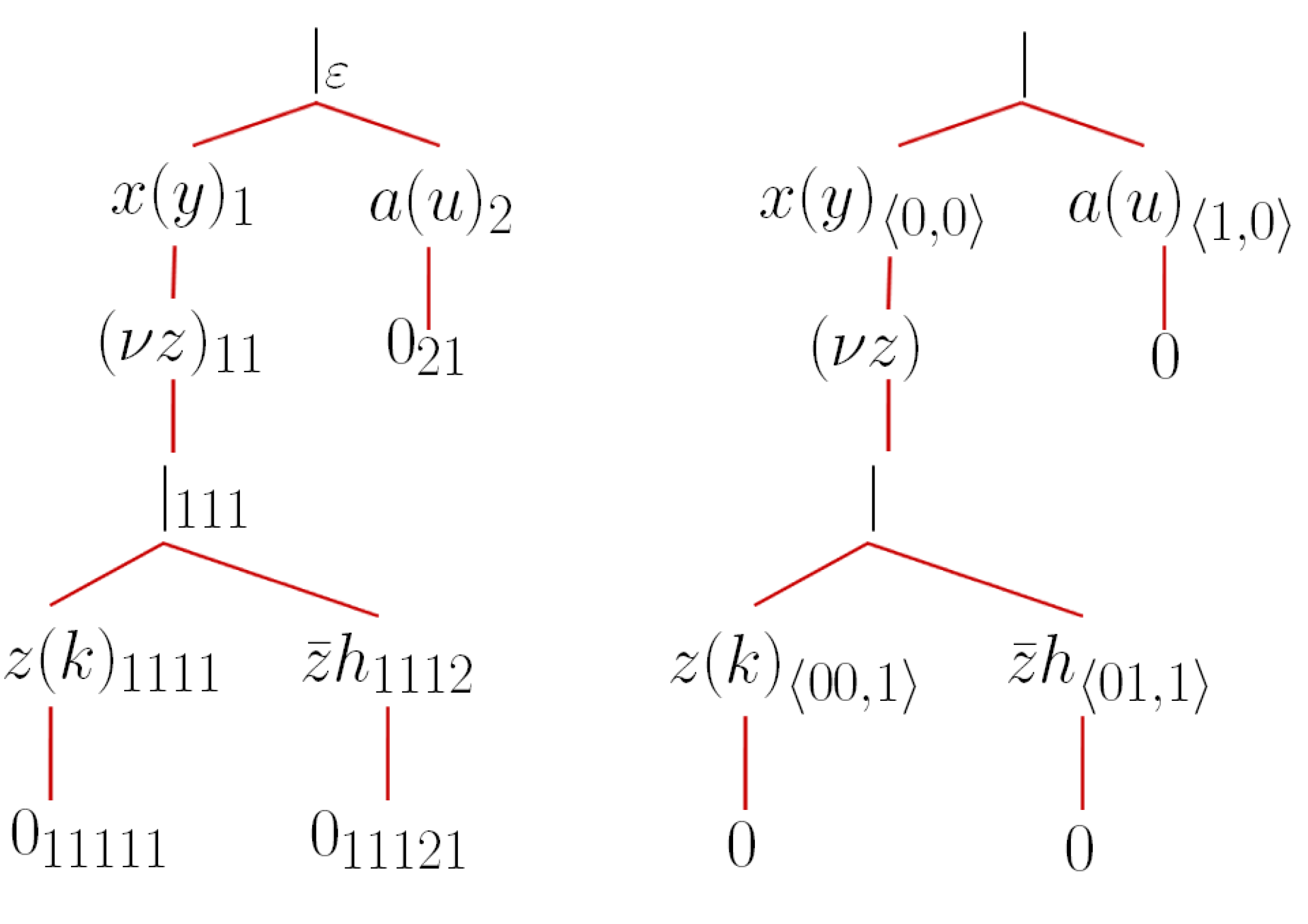}
\else
\includegraphics[width=8.75cm]{trees.eps}
\fi
\end{center}
\caption{Tree-representation of labeled terms.}\label{fig1}
\end{figure}
The representation of both labeled terms in the usual linear syntax is given in Example \ref{esa2}.
\end{exa}

\subsection{Testing with explicit fairness vs. fair-testing}
The  labeling method allows defining {\it weak-} and {\it strong-fair} computations. 
Using these notions, we adapt {\it must}-testing semantics \cite{BD95} to obtain what we call 
{\it weak-fair must}-testing semantics and {\it strong-fair must}-testing semantics. Then we compare these two 
`{\it fair}'-testing semantics with the {\it fair}-testing \cite{RV07,NC95}, that does not need 
any labeling of actions, and with the standard  {\it must}-testing. 
This comparison reveals the expressiveness of the various testing semantics we consider. 
In particular:
\begin{enumerate}[$\bullet$]
\item we show  that  {\it weak-fair must} testing  is strictly stronger than {\it strong-fair must} testing, 
\item we show  that \emph{must}-testing is strictly stronger than {\it weak-fair must} testing, 
\item we prove  that 
{\it strong-fair must} testing is strictly stronger than {\it fair}-testing, 
\item we prove that {\it strong-fair} and {\it weak-fair} {\it must}-testing  
cannot be characterized by a notion based on the transition tree, like  {\it fair}-testing.
\end{enumerate}

\subsection{Roadmap of the paper}
The rest of the paper is organized as follows. Section
\ref{piandasypi} recalls the definition of  the $\pi$-calculus. Section \ref{testing}
recalls the definition of the  {\it must}-testing   and the {\it fair}-testing semantics.
Section \ref{lablang} shows the labeling method
and its main properties. {\it Weak--fair must}- and {\it strong-fair must}-testing semantics are
defined in Section \ref{swf} and compared in Section \ref{core}.
Finally, in Section \ref{omegafair}  we investigate why strong  and
weak fairness notions are not enough to characterize {\it fair}-testing
semantics. Section \ref{fur} contains some concluding remarks and plans for future 
work. All the proofs  omitted in the body of the paper are in the 
appendixes.

\section{The \texorpdfstring{$\pi$}{pi}-calculus}\label{piandasypi}
We briefly recall here the basic notions about the
 (choiceless) $\pi$-calculus.
 Let ${\mathcal N}$ (ranged over by $x,y,z,\dots$) be a set of names. The
set ${\mathcal P}$ of processes (ranged over by $P,Q,R,\dots$) is
generated by the following grammar:
\[\begin{array}{rcl}
P & ::= & \:\: 0 \:\: \alt \:\: x(y).P \:\: 
\alt \:\: {\bar x}y.P \:\: \alt \:\: P\: |\: P \:\: \alt \:\: (\nu
x)P \:\: \alt \:\: !P \:\: \\
\end{array}
\]
The input prefix $y(x). P$, and the restriction $(\nu x) P$, act as
name binders for the name $x$ in $P$. The free names ${\it fn}(P)$
and  the bound names ${\it bn}(P)$ of $P$ are defined as usual. The
set of names of $P$ is defined as $n(P)={\it fn}(P)\cup {\it
bn}(P)$.

The operational semantics of processes is given via a labeled
transition system, whose states are the process themselves. The
labels (ranged over by $\mu , \gamma, \ldots$) ``correspond'' to
prefixes, input $xy$ and output ${\bar x}y$, and to the
bound  output ${\bar x}(y)$ (which models scope extrusion). If $\mu=
xy$ or $\mu = {\bar x}y$ or $\mu = {\bar x}(y)$ we define
$sub(\mu)=x$ and $obj(\mu)=y$. The functions ${\it fn}(\cdot)$, ${\it bn}(\cdot)$
and $n(\cdot)$ are extended to cope with labels as follows:
\[\begin{array}{llll}
{\it bn}(xy)  = \emptyset & {\it bn}({\bar x}(
y))= \{y\} & {\it bn}({\bar x}y)=  \emptyset
& {\it bn}(\tau)  = \emptyset\\
{\it fn}(xy) = \{x,y\} & {\it fn}({\bar x}(
y)) =  \{x\} &{\it fn}({\bar
x}y)=  \{x,y\} & {\it fn}(\tau) = \emptyset
\end{array}
\]
We take into account the early operational semantics for ${\mathcal P}$ in \cite{SW01}, as shown in  Table \ref{prima}. 
We only omit symmetric rules of Par, Com and Close for simplicity, and we assume alpha-conversion to avoid collision of free and bound names.

\begin{table}[!h]
\[\begin{array}{|c|}
\hline
\\
\makebox{Input}\:\: x(y).P \ar {xz}{} P\{z/y \}\:\\
\\
\makebox{Output}\:\: {\bar x}y.P \ar{{\bar x}y}{} P\:\:\\
\\
\quad\makebox{Open}\:\: \bigfrac{P \ar{{\bar x}y}{} P'}{(\nu y)P \ar
{{\bar x}(y)}{} P'}\: x \not = y \qquad
\makebox{Res}\:\:\bigfrac{P \ar{\mu}{} P'}{(\nu y)P \ar {\mu}
(\nu y)P'} \: y \not\in n(\mu)\quad\\
\\
\quad\makebox{Par}\:\:\bigfrac{P \ar{\mu} P'}{P \:| \:Q \ar{\mu}
P'\:| \:Q} \:\: {\it bn}(\mu)\cap {\it fn}(Q) = \emptyset\\
\\
\makebox{Com} \:\:\bigfrac{P \ar{xy}{} P', \:\:
Q \ar{{\bar x}y}{} Q'}{P \:| \:Q \ar{\tau}{} P' \:| \: Q'} \qquad
\makebox{Close}\:\:\bigfrac{P \ar{xy}{}
P', \:\: Q \ar{{\bar x}(y)}{} Q'}{P \:| \:Q
\ar{\tau}{} (\nu y)(P' \:| \: Q')}\: y\not\in fn(P)\quad\\
\\
\makebox{Rep} \quad \bigfrac{P \ar{\mu}{} P'}{!P
\ar{\mu}{} P'\:|\:!P}\\
\\
\hline\end{array}\]
\caption{Early operational semantics for ${\mathcal P}$ terms.}
\label{prima}
\end{table}

\begin{defi}\label{weaktrans}({\it Weak transitions})
Let $P$ and $Q$ be ${\mathcal P}$ processes. Then:
\begin{enumerate}[--]
\item $P \Ar{\varepsilon} Q$ iff $\exists\: P_0,..., P_n \in {\mathcal P}$, $n \geq 0$, s.t. $
P=P_0 \ar{\tau }...\ar{\tau} P_n =Q \: ;$
\medskip

\item $P \Ar{\mu} Q$ iff $\exists\: P_1, P_2 \in {\mathcal
P}$ s.t. $P \Ar{\varepsilon } P_1 \ar{\mu }P_2
\Ar{\varepsilon} Q \: .$
\end{enumerate}
\end{defi}

\begin{notat}
 For convenience, we write $x(y)$ and ${\bar x}y$ instead of $x(y).0$ and ${\bar x}y.0$,
respectively. Furthermore, we write $P \ar{\mu}$ (respectively $P \Ar{\mu}$) to mean that there
exists $P'$ such that $P \ar{\mu}P'$ (respectively $P \Ar{\mu} P'$) and we write
$P\Ar{\varepsilon}\ar{\mu}$ to mean that there are $P'$ and $Q$ such
that $P \Ar{\varepsilon}P'$ and $P' \ar{\mu}Q$.
\end{notat}

\section{Testing semantics}\label{testing}

In this section we briefly summarize the basic definitions behind
the testing machinery for the $\pi$-calculus.

\begin{defi}\label{observers}({\it Observers})
\begin{enumerate}[--]
\item Let $\omega\not\in {\mathcal N}$. $\omega$ denotes a special action used to report success.  
By convention ${\it fn}(\omega)={\it
bn}(\omega)=\emptyset$.
\medskip

\item The set ${\mathcal O}$  (ranged over by $o,o',o'',\ldots$)
of observers is defined like ${\mathcal P}$, where the grammar
is extended with the production $P::= \omega . P$.
\medskip

\item The operational semantics of ${\mathcal P}$ is extended to
${\mathcal  O}$ by adding $\omega . P \ar{\omega}{}P \: .$
\end{enumerate}
\end{defi}

\begin{defi}\label{experiments}({\it Experiments}) The set of experiments over ${\mathcal P}$ is defined as
\[{\mathcal E} = \{\:(P \: | \: o) \:
  | \:\: P \in {\mathcal P}\mbox{, } o \in {\mathcal O} \}
  \]
\end{defi}

\begin{defi}\label{maxcomp}({\it Maximal Computations})
Given $P \in {\mathcal P}$ { and } $o \in {\mathcal O}$, a maximal
computation from $P\: |\: o$ is either an infinite sequence of the
form
\[P\: |\: o = T_0 \ar{\tau }  T_1
 \ar{\tau }
 T_2  \ar{\tau } \ldots
\]
or a finite sequence of the form
\[P\: |\: o = T_0
\ar{\tau }  T_1  \ar{\tau } \ldots \ar{\tau }  T_n \not \ar{\tau}.
\]
\end{defi}

We are now ready to define {\it must}-  and {\it fair}-testing semantics.

\begin{defi}\label{mmf}({\it Must- and Fair-Testing Semantics})
Given a process $P \in {\mathcal P}$ and an observer $o \in {\mathcal O}$,
define:
\begin{enumerate}[--]
\item $P \: \must\: o$ if and only if {\it for every} maximal
computation from $P\: |\: o$
\[P\: |\: o = T_0  \ar{\tau }  T_1  \ar{\tau } \ldots
\ar{\tau}T_i  \: [ \ar {\tau } \ldots] 
\]
there exists $i \geq 0$ such that $T_i  \ar{\omega}$;

\item $P \fair o$ if and only if {\it for every} maximal computation from $P\: |\: o$
\[P\: |\: o = T_0  \ar{\tau }  T_1 \ar{\tau } \ldots \ar{\tau}T_i \:
[\ar{\tau } \ldots]
\]
we have $T_i\Ar{\omega}$, for every $i\geq 0$.
\end{enumerate}
\end{defi}

\section{A labeled version of the \texorpdfstring{$\pi$}{pi}-calculus} \label{lablang}
In order to deal with the notion of fairness of actions \cite{CS84},  we first need to introduce a labeling method. 
Consider  the following term: $$P=\bar a \:\: |\:\: !a.\bar a\:\:|\:\:\bar a.$$ 
Notice that every maximal computation 
from $P$ is always of the form  $$ P\rid P \rid P\rid \ldots$$ However, without labels we would not be able to distinguish 
fair computations from  unfair ones, since we
do not know which  $\bar a$ synchronizes with $!a.\bar a$ and makes progress at each step. 
So, we need to be able to refer unambiguously to individual
actions and to monitor them along any computation.
\subsection{The idea behind the labeling method}
A `reasonable' labeling method, independently from the 
choice of the labels domain, has to provide {\it unicity} (e.g. no label occurs more than
once in a labeled term), {\it disappearance} (e.g. a label disappears only when the corresponding 
action is performed) and {\it persistence} (e.g. once a label disappears, it does not appear in the
computation anymore).

The labeling method can be more or less informative, in the sense 
that the degree of information about the structure of terms (static information) and about the 
computation history (dynamic information) can vary. 
For our purpose we find useful to adopt a labeling method which is rather informative and keeps separate the  static and dynamic aspects.

\begin{defi}\label{lb}({\it Ground Labeled ${\mathcal P}$}) We define ${\mathcal P}^e_{\!{\it gr}}$ as the language generated by
the following grammar:
\[E::= 0\:\alt\: \mu_{\langle s,n\rangle}.E\:\: \alt
\: (\nu x)E\: \alt \:\: E\:|\:E \: \alt \:\: !_{\langle
s,n\rangle}P
\]
where $s\in\{0,1\}^*,\: n\in \N$,
$P\in {\mathcal P}$ and the prefix $\mu$ is of the form $x(y)$ or ${\bar x}y$.
\end{defi}

Obviously, ${\mathcal P}^e_{\!{\it gr}}$ also contains labeled terms
in which the labels do not respect the structure and/or the execution
order. To avoid this problem, we restrict the labeled language to
those terms which are {\it well-formed}. The {\it well-formedness}
predicate $\mathit{wf}(\cdot)$ (Table \ref{wf2}), allows us to obtain
a well-defined labeling method; it is defined by using a binary
relation $\Re$ over sets of labels, which checks the absence of label
conflicts in the parallel composition, and a labeling function
$L_{\langle s,n\rangle}(\cdot)$, where $s\in\{0,1\}^*$ and $n\in \N$,
which allows us to avoid label conflicts in the prefix composition.

First, we define $\Re$: if $L_0$ and $L_1$ are sets of labels,
$L_0\:\Re\: L_1$ holds if and only if for every $\langle
s_0,n_0\rangle\in L_0$ and $\langle s_1,n_1\rangle\in L_1$, the first
elements of the labels, $s_0$ and $s_1$, are not related w.r.t.  the
usual prefix relation between strings. Formally:

\begin{defi}\label{labelsetrelation}\hfil
\begin{enumerate}[1.]
\item  Given two strings  $s_0,\: s_1\in\{0,1\}^*$, we write $s_0\sqsubseteq s_1$ if 
and only if $s_0$ is a prefix of $s_1$, i.e. $s_1 = s_0 \alpha $ for some $\alpha\in \{0,1\}^*$; 
\item Given  $L_0,\: L_1\subseteq(\{0,1\}^*\times
\N)$, we write $L_0\:\Re\: L_1$ if and only if $\forall \langle
s_0,n_0\rangle\in L_0$. \, $\forall \langle s_1,n_1\rangle\in L_1$.
$s_0\not\sqsubseteq s_1$ and $s_1\not\sqsubseteq s_0$.
\end{enumerate}
\end{defi}

\begin{rem}
From Definition \ref{labelsetrelation}, it follows immediately that 
$$L_0\:\Re\: L_1\:\mbox{ implies }\:\forall\langle s_0,n_0\rangle\in L_0. \: \forall \langle 
s_1,n_1\rangle\in L_1.\:\langle s_0,n_0\rangle\not=\langle s_1,n_1\rangle.$$
\end{rem}

Then, the labeling function $L_{\langle s,n\rangle}(\cdot)$ is defined following inductively 
the ${\mathcal P}$ terms operational structure.  

\begin{defi}\label{labellingfunction} Let $P \in {\mathcal P}$.
Define $L_{\langle s,n\rangle}(P)$, where $\langle s, n\rangle\in(\{0,1\}^*\times\N)$, as in Table \ref{lf}.

\begin{table}[!h]
\[\begin{array}{|lll|}
\hline
\mbox{ } & \mbox{ } & \mbox{ } \\
L_{\langle s,n\rangle}(0) & = & 0\hfill\\
\mbox{ } & \mbox{ } & \mbox{ } \\
 L_{\langle s,n\rangle}(\mu.P) & = &   \mu_{\langle s,n\rangle}.L_{\langle s,n+1\rangle}(P) 
\\
\mbox{ } & \mbox{ } & \mbox{ } \\
L_{\langle s,n\rangle}(P_0\: |\: P_1) & = &  L_{\langle s0,n\rangle}(P_0)\: |\: L_{\langle
s1,n\rangle}(P_1)\hfill\\
\mbox{ } & \mbox{ } & \mbox{ } \\
 L_{\langle s,n\rangle}((\nu x)P) & = &  (\nu x) L_{\langle s,n\rangle}(P) \hfill\\
 \mbox{ } & \mbox{ } & \mbox{ } \\
L_{\langle s,n\rangle}(!P) & = &  !_{\langle s,n\rangle}P \hfill\\
\mbox{ } & \mbox{ } & \mbox{ } \\
\hline
\end{array}\]
\caption{Labeling function $L_{\langle s,n\rangle}(.)$.} \label{lf}
\end{table}
\end{defi}

We will use the relation $\Re$ in combination with the function $\topp(\cdot)$, defined in Table \ref{rad},  
which gives  the top-level label set of a labeled term.
In the same table we define also the function  $lab(\cdot)$, which returns the whole set of labels, and which will be useful later.

\begin{table}[!h]
\[\begin{array}{|lll|}
\hline
\mbox{ } & \mbox{ } & \mbox{ } \\
E=0\!:&\topp(E) = \emptyset &\:\:\: lab(E)=\emptyset\\
\mbox{ } & \mbox{ } & \mbox{ } \\
E=\mu_{\langle s,n\rangle}.E'\!:&\topp(E)=\{\langle s,n\rangle\} & \:\:\:
lab(E)=\{\langle s,n\rangle\}\cup lab(E')\\
\mbox{ } & \mbox{ } & \mbox{ } \\
E=(\nu x)E'\!:&\topp(E)=\topp(E') &  \:\:\:lab(E)=lab(E')\\
\mbox{ } & \mbox{ } & \mbox{ } \\
E=E_0|E_1\!:&\topp(E)=\topp(E_0)\cup\topp(E_1) &\:\:\:  lab(E)=lab(E_0)\cup lab(E_1)\\
\mbox{ } & \mbox{ } & \mbox{ } \\
E=!_{\langle s,n\rangle}P\!:&\topp(E)=\{\langle s,n\rangle\}  & \:\:\:lab(E)=\{\langle s,n\rangle\}\\
\mbox{ } & \mbox{ } & \mbox{ } \\
\hline
\end{array}\]
\caption{Function $\topp(\cdot)$ and $lab(\cdot)$.} \label{rad}
\end{table}

\begin{rem}\label{f} 
From the definitions in Table \ref{rad},  we have that
$$\forall E\in{\mathcal P}^e_{\!{\it gr}}.\:\topp(E)\subseteq lab(E)$$
\end{rem}

Finally, Table \ref{wf2} defines formally the well-formedness predicate $\mathit{wf}(\cdot)$. Note that 
we use $\Re$ to check the lack of conflict, between labels in parallel components, 
at the top-level only. This constraint will turn out to be sufficient. In fact, in Lemma \ref{add1} in the appendix it is proved that 
$$\topp(E_0)\:\Re\: \topp(E_1)\:\mbox{ implies }\: lab(E_0)\:\Re\: lab(E_1).$$

\begin{table}[!h]
\[\begin{array}{|c|}
\hline
\\
\quad\makebox{Nil} \quad \bigfrac{}{\mathit{wf}(0)}\quad\quad
\makebox{Pref} \quad \bigfrac{\mu.P\in {\mathcal P}}{\mathit{wf}(L_{\langle s,n\rangle}(\mu.P))}\quad\\
\\
\quad\makebox{Par}  \quad \bigfrac{\mathit{wf}(E_0),\quad \mathit{wf}(E_1),\quad \topp(E_0)
\:\Re \:\topp(E_1)}{\mathit{wf}(E_0\: |\: E_1)}\quad\\
\\
\makebox{Res} \quad \bigfrac{\mathit{wf}(E)}{\mathit{wf}((\nu x)E)}\quad
\quad\makebox{Rep} \quad \bigfrac{P\in{\mathcal P}}{\mathit{wf}(!_{\langle s,n\rangle}P)}\\
\\
\hline
\end{array}\]
\caption{Well formed terms.} \label{wf2}
\end{table}

Now we are ready to define  the set of labeled ${\mathcal P}$-calculus terms, denoting it by 
${\mathcal P}^e$. 

\begin{defi}\label{labelledlanguage2}  The labeled ${\mathcal P}$-calculus, denoted by ${\mathcal P}^e$, is the set
\[\{E\in {\mathcal P}^e_{\!{\it gr}}\: |\: \mathit{wf}(E)\}\]
\end{defi}

It would be possible to defined well-formed terms without explicitly relying on the labeling function: 
for example, defining an ordering relation between labels to characterize well-formedness of prefixing. 
However, our aim is to keep separated static and dynamic informations. More in detail, 
${\mathcal P}^e$ contains all the well-formed processes of the form `$L_{\langle s,n\rangle}(P)$' (Lemma 
\ref{starstar}). However, the operational semantics of ${\mathcal P}^e$, introduced in the following, 
does not preserve the `$L_{\langle s,n\rangle}(P)$' format: for this reason, the $\mathit{wf}(.)$ predicate is defined 
in order to ensure the closure of ${\mathcal P}^e$ w.r.t  $\rid$.

\begin{exa}\label{esa2} Consider again the term 
$S\: =\:x(y).((\nu z)(z(k).0\:\: |\:\: {\bar z}h.0))\:\: |\:\: a(u).0\:$ of Example \ref{esa1}.
In the approach of \cite{CS84, CS87}, the  labeling of $S$ would give the term 
\[ x(y) _{1} . ((\nu z)_{11}(z(k)_{1111} .0 _{11111} \:|_{111} \: 
{\bar z}h_{1112} .0_{11121})) \: |_{\varepsilon} \: a(u)_{2} .0_{21}.
\]
In our approach, the  labeling of $S$ is the term 
\[x(y)_{\langle 0,0\rangle}.((\nu z)(z(k)_{\langle 00,1\rangle}.0\:\:|\:\:
{\bar z}h_{\langle 01,1\rangle}.0))\:\: |\:\: a(u)_{\langle 1,0
  \rangle}.0.
\]
\end{exa}

\subsection{Some properties of the labeled \texorpdfstring{$\pi$}{pi}-calculus}\label{plp}

The operational semantics of  ${\mathcal P}^e$ is similar to the one in Table \ref{prima};
we simply ignore labels in order to derive a transition. The only rule that needs attention is the one for 
processes in the scope of the replication operator, since the unfolding generates new parallel processes and
we must ensure unicity, disappearance and persistence of labels. We use the dynamic labeling  described in Table \ref{bang}.

\begin{table}[!h]
\[\begin{array}{|c|}
\hline
\\
\makebox{Rep}\quad \bigfrac{P \ar{\mu}{} P'}{!_{\langle
s,n\rangle}P \ar{\mu}{} L_{\langle
s0,n+1\rangle}(P')\: |\:!_{\langle s1,n+1\rangle} P}
\quad\\
\\
\hline
\end{array}\]
\caption{Replication Rule in ${\mathcal P}^e$.} \label{bang}
\end{table}

${\mathcal P}^e$ is trivially closed w.r.t. renaming,  since a renaming 
does not change labels. It follows that the language is closed w.r.t.  $\rid$. 

Next result  states the main properties which make our labeling method `reasonable': 

\begin{thm}\label{a} Let $E\in {\mathcal P}^e$.
\begin{enumerate}[\em 1.]
\item {\it (Unicity)} No label $\langle s,n\rangle$ occurs more than once in
$E$;
\item {\it (Disappearance)} If $E\ar{\mu}{} E'$ then $\exists\langle
s,n\rangle\in lab(E).\: \langle s,n\rangle\not\in lab(E')$;

\item {\it (Persistence)} $\forall k\geq 1.$  $E_0\ar{\mu_0}{} E_1\ar{\mu_1}{}
E_2\ar{\mu_2}{}\dots\ar{\mu_{k-1}}{} E_k$, if $\langle s,n\rangle\in
lab(E_0)\cap lab(E_k)$ then $\langle s,n\rangle\in lab(E_i)$ for any $i\in[1..(k-1)]$.
\end{enumerate}
\end{thm}

\proof\hfil
\begin{enumerate}[(1)]
\item By induction on the structure of $E$.
\begin{enumerate}[--]
\item $E=0$: then $lab(0)=\emptyset$.
\item $E=L_{\langle s,n\rangle}(\mu.P')$: then $lab(E)=\{\langle
s,n\rangle\}\cup lab(L_{\langle s,n+1\rangle}(P'))$. By Lemma \ref{starstar},  
$\mathit{wf}(L_{\langle s,n+1\rangle}(P'))$, i.e. $L_{\langle s,n+1\rangle}(P')\in {\mathcal P}^e$ and, 
by induction hypothesis, for every
$ \langle s',n'\rangle\in lab(L_{\langle
s,n+1\rangle}(P'))$, $\langle s',n'\rangle$ does not occur more
than once in $lab(L_{\langle s,n+1\rangle}(P'))$.  By Lemma \ref{star},  
$\forall \langle s',n'\rangle\in lab(L_{\langle
s,n+1\rangle}(P'))$. $s\sqsubseteq s'$ and $n+1\leq n'$. Hence $\langle s,n
\rangle\not \in lab(L_{\langle s,n+1\rangle}(P'))$.
\item $E=(E_0\: |\: E_1)$: by definition, $\forall i\in\{ 0,1\}. \: \mathit{wf}(E_i)$ holds, implying 
$E_i\in {\mathcal P}^e$, and  $lab(E)= \bigcup_i lab(E_i)$. By
induction hypothesis, for every $ i\in\{ 0, 1\}$ and every $\langle
s_i,n_i\rangle\in lab(E_i)$, $\langle s_i,n_i\rangle$ does not
occur more than once in $lab(E_i)$. By Lemma \ref{add1}, 
$\topp(E_0)\:\Re\: \topp(E_1)$ implies 
$lab(E_0)\:\Re\: lab(E_1)$,  i.e. $\forall
\langle s_0,n_0\rangle\in lab(E_0). \: \forall \langle
s_1,n_1\rangle\in$ $lab(E_1)$.  $\langle s_0,n_0\rangle\not=\langle
s_1,n_1\rangle$. Hence,  for every $ i\in\{0,1\}$ and every 
$ \langle s_i,n_i\rangle\in lab(E_i)$, $\langle
s_i,n_i\rangle$ does not occur more than once in $lab(E)$.

\item Cases $E=(\nu x)E'$ and $E=!_{\langle s,n\rangle}P$
can be proved similarly.
\end{enumerate}
\item By Remark \ref{f} and Lemma \ref{add2}, it suffices to prove that $$E\ar{\mu}{}
E'\mbox{ implies }\exists\langle s,n\rangle\in \topp(E).\:\langle
s,n\rangle\not\in \topp(E').$$ In fact,  $\langle s,n\rangle\in \topp(E)$ implies 
$\langle s,n\rangle\in lab(E)$ (by Remark \ref{f}), and $\langle
s,n\rangle\not\in \topp(E')$ implies $\langle
s,n\rangle\not\in lab(E')$ (by Lemma \ref{add2}).

 By induction on the depth of $E\ar{\mu}{}E'$.
\begin{enumerate}[--]
\item {\it Rule Input/Output}: $E=L_{\langle
s,n\rangle}(\mu.P)\ar{\mu}{}E'=L_{\langle s,n+1\rangle}(P')$
(either $P' = P$ or $P' = P\{z/y\}$). $\topp(L_{\langle
s,n\rangle}(\mu.P))=\{\langle s,n\rangle\}$ and, by Lemma \ref{star} on 
$L_{\langle s,n+1\rangle}(P')$,  $\forall \langle r',m'\rangle\in \topp(L_{\langle
s,n+1\rangle}(P'))\subseteq lab(L_{\langle
s,n+1\rangle}(P'))$. $s\sqsubseteq r'$ and $n+1\leq m'$. Hence 
$\langle s,n\rangle\in\topp(E)$ and 
$\langle s,n\rangle\not \in \topp(L_{\langle s,n+1\rangle}(P'))$.

\item {\it Rule Par}: $E=(E_0 \:| \:E_1) \ar{\mu} (E_0'\:| \:E_1)$,
where $bn(\mu)\cap fn(E_1) = \emptyset$ and  $E_0 \ar{\mu} E_0'$.
By induction hypothesis, $\exists\langle r_0,m_0\rangle\in \topp(E_0).\:
\langle r_0,m_0\rangle\not\in  \topp(E_0')$. Since  
$\topp(E_0)\:\Re \:\topp(E_1)$ holds, then  $\{\langle r_0,m_0\rangle\}\Re \topp(E_1)$, i.e. 
$\langle r_0,m_0\rangle\not\in \topp(E_1)$. 
We conclude that  $\langle r_0,m_0\rangle\not\in \topp(E_0'\:| \:E_1)$.

\item {\it Rule Com}: $E=(E_0 \:| \:E_1) \ar{\tau} (E_0'\:| \:E_1')$,
where $E_0 \ar{xy} E_0'$ and $E_1 \ar{{\bar x}y}{} E_1'$. 
By induction hypothesis, $\exists\langle r_0,m_0\rangle\in \topp(E_0).\:
\langle r_0,m_0\rangle\not\in \topp(E_0')$  and 
$\exists\langle r_1,m_1\rangle\in \topp(E_1).\:
\langle r_1,m_1\rangle\not\in  \topp(E_1')$. 

Consider $\langle r_0,m_0\rangle$  
(case $\langle r_1,m_1\rangle$ is symmetric). 
Since  $\mathit{wf}(E_0\:| \:E_1)$, then we have $\topp(E_0)\:\Re \:\topp(E_1)$. 
This implies  $\{\langle r_0, m_0\rangle\}\Re \topp(E_1)$.

By  item (3) of Lemma \ref{3-l5555}  on $E_1 \ar{{\bar x}y}{} E_1'$, 
$\forall \langle r_1', m_1'\rangle\in \topp(E_1')$. 
$\exists \langle r', m'\rangle\in \topp(E_1).$ $r'\sqsubseteq r_1'$ and $m'\leq m_1'$.    
By Lemma \ref{cat1}, it follows that $\{\langle r_0,m_0\rangle\}\:\Re \:\topp(E_1')$, and therefore
$\langle r_0,m_0\rangle\not\in \topp(E_1')$. 
We can conclude that $\langle r_0,m_0\rangle\in \topp(E_0\:| \:E_1)$ and 
$\langle r_0,m_0\rangle\not\in \topp(E_0')\cup \topp(E_1')= \topp(E_0'\:| \:E_1')$.    

\item {\it Rule Open/Res/Close/Rep}: These cases can be proved
similarly.
\end{enumerate}

\item  In \cite{CS87} (Lemma 8.8), the analogous property is only proved for $k=2$. However, the general case cannot be obtained by induction, since 
the reasoning for the case   $k=2$ does not contain the essential elements to prove the inductive step. Differently from \cite{CS87}, we prove the property in the general case. We proceed as follows.

By contradiction, let $i\in[1..(k-1)]$ be the least index such that $\langle s,n\rangle\not\in lab(E_i)$ and 
let $j\in[(i+1)..k]$ be the least index such that $\langle s,n\rangle\in lab(E_j)$. By the minimality of $i$, we can apply Lemma \ref{add3} and we obtain that 
$\langle s,n\rangle\in \topp(E_{i-1})$. By item (2) of 
Lemma \ref{3-l5555} on $E_j$, $\exists \langle r_j, m_j\rangle\in \topp(E_j).$ $r_j\sqsubseteq s$ and $m_j\leq n$. 
By  item (3) of Lemma \ref{3-l5555}  on $E_c \ar{\mu_c}{} E_{c+1}$ for any $c\in[(i-1)..(j-1)]$, $\exists \langle r_c, m_c\rangle\in 
\topp(E_c).$ $r_c\sqsubseteq r_{c+1}$ and $m_c\leq m_{c+1}$. It follows that $\exists \langle r_{i-1}, m_{i-1}\rangle \in 
\topp(E_{i-1}).$ $r_{i-1}\sqsubseteq s$ and $m_{i-1}\leq n$. 
\begin{enumerate}[--]
\item In the case $\langle r_{i-1}, m_{i-1}\rangle$ and 
$\langle s,n\rangle$ are distinct labels: we contradict item (1) of Lemma \ref{3-l5555}. 
\item In the case $\langle r_{i-1}, m_{i-1}\rangle=\langle s,n\rangle$: it follows that 
$\forall c\in((i-1)..(j-1)].$ $\exists \langle r_c, m_c\rangle\in 
\topp(E_c).$ $s\sqsubseteq r_c\sqsubseteq s$ and $n\leq m_c\leq n$, i.e. $s= r_c$ and $n=m_c$, contradicting 
that $\langle s, n\rangle\not\in lab(E_c)$.\qed
\end{enumerate}
\end{enumerate}

\begin{rem} The disappearance property states that a label {\em
    disappears} when the corresponding action is performed. On the
  other hand, the persistence ensures a {\em complete disappearance}
  of a label, once the corresponding action is performed.  In fact, it
  is clear that for $E_0\ar{\mu_0}{}
  E_1\ar{\mu_1}{}...\ar{\mu_{k-1}}{} E_k$ with $\langle s,n\rangle\in
  \!lab(E_0)\cap lab(E_k)$ the existance of some $h\in[1..(k-1)]$
  satisfying $\langle s,n\rangle\not\in lab(E_h)$ would contradict
  item (3).
\end{rem}

As expected, the labeled language is a {\it conservative extension} of
the unlabeled one.  To prove the statement, we have to formally define
the ${\mathcal P}$ process that is obtained by deleting all the labels
appearing within a labeled term.

\begin{defi}\label{unlab} Let $E\in {\mathcal P}^e$. Define $Unl(E)$ as
the ${\mathcal P}$ process obtained by removing all the labels in $E$.
It can be defined by induction as in Table \ref{u}.
\begin{table}[!h]
\[\begin{array}{|lll|}
\hline
\mbox{ } & \mbox{ } & \mbox{ } \\
Unl(0) & = & 0\hfill\\
\mbox{ } & \mbox{ } & \mbox{ } \\
Unl(\mu_{\langle s,n\rangle}.E) & = &  \mu.Unl(E)\\
\mbox{ } & \mbox{ } & \mbox{ } \\
Unl(E_0\: |\: E_1) & = &  Unl(E_0)\: |\: Unl(E_1)\hfill\\
\mbox{ } & \mbox{ } & \mbox{ } \\
Unl((\nu x)E) & = &  (\nu x)Unl(E) \hfill\\
\mbox{ } & \mbox{ } & \mbox{ } \\
Unl(!_{\langle s,n\rangle}P)  & = &  !P\hfill\\
\mbox{ } & \mbox{ } & \mbox{ } \\
\hline
\end{array}\]
\caption{Function $Unl(\cdot)$.} \label{u}
\end{table}
\end{defi}

The conservative property of the labeled extension is expressed by the following 
lemma, which can be proved by induction on the depth of
$E\ar{\mu}{}E'$ (item (1)) and $Unl(E) \ar{\mu}{} P'$ (item (2)).
\begin{prop}\label{b} Let $E\in {\mathcal P}^e$.
\begin{enumerate}[\em 1.]
\item $E\ar{\mu}{}E'$ implies $Unl(E) \ar{\mu}{} Unl(E')$;

\item $Unl(E) \ar{\mu}{} P'$ implies $\exists E' \in {\mathcal P}^e.$ 
$E\ar{\mu}{}E'$ and $Unl(E')=P'$.\qed
\end{enumerate}
\end{prop}

\section{Strong and weak fairness of actions}\label{swf}

The  labeling method proposed in the previous section   can be  extended in a natural way 
over the observers, adding $B::= \omega.B$ in the grammar of 
${\mathcal P}^e_{\!{\it gr}}$, $\omega.o\ar{\omega}{}o$ in the operational 
semantics and extending the functions $L_{\langle s,n\rangle}$, $\topp(\cdot)$, $lab(\cdot)$, $Unl(\cdot)$ 
and the predicate $\mathit{wf}(\cdot)$ as shown in Table \ref{omegalab}. No label is associated to $\omega$  
since we do not need to distinguish $\omega$ occurrences\footnote{$E\ar{\omega}$   
whenever an {\em arbitrary} occurrence of $\omega$ is at the top level in $E$.}.  
\par\smallskip
\begin{table}[thb]
\[\begin{array}{|lllll|}
\hline
\mbox{ } & \mbox{ } & \mbox{ } & \mbox{ } & \mbox{ }\\
\qquad\makebox{($L_{\langle s,n\rangle}/Unl$)}\quad & L_{\langle s,n\rangle}(\omega.o)\: & = &\: 
Unl (L_{\langle s,n\rangle}(\omega.o)) \quad & =
\quad\omega.o\qquad\\
\mbox{ } & \mbox{ } & \mbox{ }  & \mbox{ }  & \mbox{ }\\ 
\qquad\makebox{($\topp/lab$)} \quad &\topp(\omega.o) \: & = &\:  lab(\omega.o)\quad & =\quad\emptyset\qquad\\
\mbox{ } & \mbox{ } & \mbox{ }  & \mbox{ }  & \mbox{ }\\ 
\qquad\makebox{($\mathit{wf}$)}\quad & \bigfrac{\omega.o\in {\mathcal O}}{\mathit{wf}(\omega.o)} & \mbox{ } & \mbox{ } & \mbox{ }\qquad\\
\mbox{ } & \mbox{ } & \mbox{ } & \mbox{ } & \mbox{ }\\
\hline
\end{array}\]
\caption{Labeling method extension over observers.} \label{omegalab}
\end{table}

In the following, ${\mathcal O}^e$ (ranged over by $\rho,\rho',..$)
denotes the set of labeled observers and ${\mathcal E}^e$ denotes the
set of labeled experiments over ${\mathcal P}^e$, as expected.

The definition of {\it live label} is crucial in the notion of
fairness.  Given a labeled experiment $S\in {\mathcal E}^e$, a {\it
  live label} is a label associated to a top-level action which can
immediately be performed, i.e. an input/output prefix able to
synchronize.  Table \ref{live} defines the live labels of a labeled
experiment $S\in {\mathcal E}^e$, according to the labeling method
proposed in Section \ref{lablang}. Informally, Table \ref{live} is a
rephrasing of operational rules: even if live labels cannot be
directly defined in term of transitions, deductions of live predicate
mime the proof for a derivation.  As a consequence, $\omega$ is not
live, since a complementary action ($\overline \omega$) does not
exist.  Given a labeled experiment $S$, the set of $S$ live labels is
denoted by $Ll(S)$.

\begin{defi}\label{livelabels} Let $S\in {\mathcal E}^e$,
let $\langle s,n\rangle\in (\{0,1\}^*\times \N)$.
\[Ll(S)=\{\langle s,n\rangle\in (\{0,1\}^*\times \N)\:|\:\: \makebox{\it
live}(\langle s,n\rangle, \tau, S)\}
\]
is the set of live labels associated to initial $\rid$ from $S$.
\end{defi}

If $S\not\!\!\rid$, then $Ll(S)=\emptyset$. 
Since $\topp(S)$ is defined as
the set of  labels appearing at the top of $S$, it follows
immediately by the definition of live actions that $Ll(S)\subseteq \topp(S)$. 
For  simplicity, labels
will be denoted in the following by $v, v_1, v_2,\ldots\in(\{0,1\}^*\times\N)$.

\begin{table}[t]
\[\begin{array}{|c|}
\hline
\\ 
\makebox{Input} \quad \bigfrac{x,y,z \in {\mathcal N} }{\makebox{\it
live}( \langle s,n\rangle, xz, x(y)_{\langle
s,n\rangle}.S)}\:\\
\\
\:\makebox{Output} \quad
\bigfrac{x,z\in {\mathcal N}}{\makebox{\it live}( \langle s,n\rangle, {\bar x}z,
{\bar x}z_{\langle s,n\rangle}.S)}\quad
\makebox{Res} \quad \bigfrac{\makebox{\it live}(\langle
s,n\rangle,\mu, S)\:\:\:y\not\in n(\mu)}{\makebox{\it
live}(\langle s,n\rangle,\mu, (\nu y)S)}\:\\
\\
\:\makebox{Open} \quad \bigfrac{\makebox{\it live}(\langle
s,n\rangle,{\bar x}y, S)\:\:\:x\not=y}{\makebox{\it live}(\langle
s,n\rangle,{\bar x}(y), (\nu y)S)}\:\:\:\:
\makebox{Rep}\quad \bigfrac{S \ar{\mu}{} S'}{\makebox{\it live}(\langle
s,n\rangle, \mu, !_{\langle s,n\rangle} S)}\:\: 
\\
\\
\makebox{Par}  \quad
\bigfrac{\makebox{\it live}(\langle s,n\rangle,\mu,
S_0)\:\:\:bn(\mu)\cap fn(S_1) = \emptyset}{\makebox{\it
live}(\langle s,n\rangle,\mu, (S_0\: |\:
S_1))}\\
\\
\makebox{Com}  \:\bigfrac{\makebox{\it live}(\langle
s,n\rangle, xy, S_0),\:\:\makebox{\it live}(\langle
r,m\rangle,{\bar x}y, S_1)}{\makebox{\it live}(\langle s,n\rangle\:,\tau, S_0\: |\: S_1),\:
 \makebox{\it live}(\langle r,m\rangle\:,\tau, S_0\: |\: S_1)}\\
 \\
\makebox{Close} \:\bigfrac{\makebox{\it live}(\langle
s,n\rangle, xy, S_0),\:\:\makebox{\it live}(\langle
r,m\rangle,{\bar x}(y), S_1), \:\:y\not \in fn(S_0)}{\makebox{\it live}(\langle s,n\rangle\:,\tau, (\nu y)(S_0\: |\: S_1)),\:
 \makebox{\it live}(\langle r,m\rangle\:,\tau, (\nu y)(S_0\: |\: S_1))}\\
 \\
\hline
\end{array}\]
\caption{Live labels.} \label{live}
\end{table}
\par

We can now formally define the strong and weak notions of fairness.
Intuitively, a {\it weak-fair} computation is a maximal computation such that no
label  becomes live and then stays live forever. 

\begin{defi}\label{weakmaxcompfair}({\it Weak-fair Computations})
Given $S\in {\mathcal E}^e$, a {\it weak-fair computation} from $S$ is a
maximal computation,
\[ S = S_0\ar{\tau } S_1\ar{\tau
}S_2\ar{\tau } \ldots \ar{\tau}S_i\:[\ar{\tau}\ldots]
\]
where $\forall v\in (\{0,1\}^*\times\N)$. $\forall i\geq 0.\:
\exists j\geq i .$  $v\not\in Ll(S_j)$.
\end{defi}

A {\it strong-fair} computation is a maximal computation such that no label
is live infinitely often. Formally, strong fairness
imposes that for every label there is some point beyond which it 
is never live.  

\begin{defi}\label{strongmaxcompfair}({\it Strong-fair Computations})
Given $S\in {\mathcal E}^e$, a {\it strong-fair computation} from $S$ is
a maximal computation,
\[ S  = S_0\ar{\tau } S_1\ar{\tau
}S_2\ar{\tau } \ldots \ar{\tau}S_i\: [\ar{\tau}\ldots]
\]
where $\forall v\in (\{0,1\}^*\times\N)$. $\exists i\geq 0 .$
$\forall j\geq i$.  $v\not\in Ll(S_j)$.
\end{defi}

Note that every finite computation is {\it strong-fair} (resp. {\it
  weak-fair}), because there is no transition $\rid$ from the end
state, which implies that there are no live labels.

Some useful results follow:

\begin{thm}\label{strongweak} $\forall S\in {\mathcal E}^e$.
\begin{enumerate}[\em1.]
\item there is always a {\it strong-fair} computation from $S$, and

\item every {\it strong-fair} computation from $S$ 
is  {\it weak-fair}, but not vice versa.
\end{enumerate}
\end{thm}

\proof\hfil
\begin{enumerate}[(1)]
\item We apply items of Lemma \ref{lemma0}.  If $S\not\ar{\tau}$, then the empty
computation is {\it strong-fair}, since $Ll(S)=\emptyset$. Otherwise,  there is
a maximal computation ${\mathcal C}$
\[S=S_0\rid S_0^1\rid..\rid S_0^{n_0}\rid S_1\:[
\rid S_1^1\rid..\rid S_1^{n_1}\rid S_2\rid\ldots]
\]
where $\forall i\geq 0. \: Ll(S_i)\cap Ll(S_{i+1})=\emptyset$ and
$\forall j\geq i. \: Ll(S_i)\cap Ll(S_j)=\emptyset$. Suppose, by
contradiction, that ${\mathcal C}$ is not {\it strong-fair}: then
there exists a label $v$ such that $\forall i\geq 0$. $\exists j\geq
i.$ $v\in Ll(\tilde S)$, where either $\tilde S=S_j$ or $\tilde
S=S_j^{k}$, contradicting the hypothesis on ${\mathcal C}$.

\item The positive result is trivial: by definition, a {\it strong-fair} computation is a special 
case of {\it weak-fair} computation. To prove
the negative result, let $S=E\: |\:\rho$, where $E= !_{v^0_1}a\: |\: (\nu b)({\bar b}_{v^0_2}\: 
|\:!_{v^0_3}b_.({\bar a}\:  |\:{\bar b}))$ and $\rho= a_{v_4}.\omega$: it is not difficult to check that there exists a  maximal
computation from $S$, along which $a_{v_4}$ is never performed.  The maximal computation 
${\mathcal C}$ we consider 
is the following one (we omit $0$ term by convenience): 
\[E\: |\:\rho= S_0\rid S_1\rid S_2\rid \ldots \rid S_i\rid\]
where $\forall j\geq 0. \: Q(v^j_{2}, v^j_{3})\: = \:(\nu b)
({\bar b}_{v^j_2}\: | \:!_{v^j_3}b.(\:{\bar a}\: \: |\:\:{\bar b}))\:$ and
\[\begin{array}{lll}
S_0= !_{v^0_1}a\: |\:Q(v^0_{2}, v^0_{3})\: |\:a_{v_4}.\omega & \mbox{ } &\qquad \ldots\\
S_1= !_{v^0_1}a\: |\: {\bar a}_{v^1_5}\: |\:Q(v^1_{2}, v^1_{3})\: |\:a_{v_4}.\omega & \mbox{ } & 
\qquad S_i= !_{v^i_1}a\: |\:Q(v^{i-1}_{2}, v^{i-1}_{3})\: |\:a_{v_4}.\omega\\
S_2= !_{v^2_1}a\: |\:Q(v^1_{2}, v^1_{3})\: |\:a_{v_4}.\omega & \mbox{ } & 
\qquad S_{i+1}= !_{v^i_1}a\: |\: {\bar a}_{v^{i+1}_5}\: |\:Q(v^{i+1}_{2}, v^{i+1}_{3})\: |\:a_{v_4}.\omega\\
S_3= !_{v^2_1}a\: |\: {\bar a}_{v^3_5}\: |\:Q_2(v^3_{2}, v^3_{3})\: |\:a_{v_4}.\omega & \mbox{ } & 
\qquad S_{i+2}= !_{v^{i+2}_1}\:a\: |\:Q(v^{i+1}_{2}, v^{i+1}_{3})\: |\:a_{v_4}.\omega\\
S_4= !_{v^4_1}a\: |\:Q(v^3_{2}, v^3_{3})\: |\:a_{v_4}.\omega & \mbox{ } & \qquad \ldots\\
\end{array}
\]
Notice that, in ${\mathcal C}$, we have $v_4\not\in Ll(S_0), v_4\in Ll(S_1), v_4\not\in Ll(S_2), v_4\in Ll(S_3),$
$\ldots, v_4\not\in Ll(S_i), v_4\in Ll(S_{i+1}), v_4\not\in Ll(S_{i+2}), \ldots$ and so on.  Moreover
for every $ v\in Ll(S_j), $ where $v\not= v_4$, there exists $k>j$ such that $v\not\in Ll(S_k)$.
I.e.,  ${\mathcal C}$ is {\it weak-fair} but it is not {\it strong-fair}.
\qed
\end{enumerate}

\section{Comparing `fair'-testing semantics}\label{core}
In this section we  consider the addition of the requirement of fairness in the definition of 
the {\it must}-testing  and investigate the resulting semantic relations. In particular, we
compare the different notions 
of fairness (the notions we introduce and the existing notion of {\it fair}-testing semantics), and 
the {\it must}-testing semantics.

Let us start  by observing that 
$P\must o$ implies $P \fair o$, but not vice versa:
it suffices to consider  the process $P=(\nu b)({\bar b}\: |\:!b.{\bar b})\: |\:{\bar a}$
and the observer $o=a.\omega$. 

Now, we define our notions of `fair' {\it must}-testing. 

\begin{defi}\label{fmust}({\it Strong/Weak-fair Must Semantics})
Let $E\in{\mathcal P}^e$ and $\rho \in {\mathcal
O}^e$. Define $E \sfmust \rho$ ($E \wfmust \rho$) if and only if {\it for every} {\it strong-fair} 
(respectively, {\it weak-fair}) computation from $(E\: |\: \rho)$
\[E\: |\: \rho = S_0 \ar{\tau }  S_1
 \ar{\tau } \ldots \ar{\tau} S_i \:[\ar{\tau }
\ldots]
\]
there exists some $i\geq 0$ such that $S_i \ar{\omega}$.
\end{defi}

The following result  states the relation between 
{\it weak-fair must}-testing   and {\it strong-fair must}-testing. It is the case that
{\it weak-fair must}-testing  implies {\it strong-fair must}-testing, but
not vice versa. In fact, any  {\it strong-fair} computation
is also  {\it weak-fair}. To prove the negative result, we consider an experiment 
with {\it weak-fair} computation in which the label prefixing $\omega$ becomes live, loses
its liveness, becomes live again, etc., without being performed:
this computation is {\it weak-fair} by definition and unsuccessful. Notice
that  this label should be always performed in a {\it strong-fair}
computation, determining the success of it. 

\begin{prop}\label{risultatoclou} $\forall E \in {\mathcal P}^e.$ $\forall\rho \in {\mathcal O}^e$. 
\begin{center}
  $E\wfmust \rho$ implies $E \sfmust \rho$, but not vice versa.
\end{center}
\end{prop}
\proof For the positive part, suppose,  by contradiction, that there exists a 
{\it strong-fair} computation ${\mathcal C}$
$$E\: |\: \rho = S_0 \ar{\tau } S_1
 \ar{\tau } \ldots  \ar{\tau}S_i \:[\ar{\tau }
\ldots] $$
such that $\forall i\geq 0. \: S_i\not\ar{\omega}$. Since a {\it strong-fair}
computation is {\it weak-fair} too, then ${\mathcal C}$ is {\it weak-fair}. It follows that $E\notwfmust \rho$,
thus contradicting the hypothesis.

We now prove the negative result. Consider  again $E= !_{v^0_1}a\: |\:Q(v^0_{2}, v^0_{3})$
and $\rho= a_{v_4}.\omega$, where $Q(v^j_{2}, v^j_{3})= (\nu b)({\bar b}_{v^j_2}\: | !_{v^j_3}b.({\bar a}\:  |\:{\bar b})).$

Notice that the computation  proposed in the proof of item (2) of Theorem \ref{strongweak}, where  $v_4\not\in Ll(S_0),
v_4\in Ll(S_1), v_4\not\in Ll(S_2)$, $v_4\in Ll(S_3),..,
v_4\not\in Ll(S_j), v_4\in Ll(S_{j+1})$,  $v_4\not\in Ll(S_{j+2})$ etc.,    
is unsuccessful: in fact,  $v_4$ loses its liveness even if $a_{v_4}$ is not performed. In such a case
$\forall j\geq 0. \: S_{j}\not\ar{\omega}$. It follows that  $E \notwfmust \rho$.

To prove that $E\sfmust \rho$ holds, it suffices to notice that  for every $j\geq 0$ and every $v^j_{2}, v^j_{3}\in (\{0,1\}^*\times\N)$,
\begin{enumerate}[(1)]
\item  $Q(v^j_{2}, v^j_{3})\rid  {\bar a}_{v^{j+1}_5}\: 
  |\:Q_2(v^{j+1}_{2}, v^{j+1}_{3})$,  i.e. 
  $Q(v^j_{2}, v^j_{3})$ can perform infinite $\rid$ sequences;

\item for every $T\in{\mathcal E}^e$, every $\rid$ from $(Q(v^j_{2}, 
  v^j_{3})\:|\:T)$  does not follow from a synchronization (either
  Rule Com or Close) between $Q(v^j_{2}, v^j_{3})$ and $T$;

\item  for every maximal computation ${\mathcal C'}$ from $E\: |\:\rho$
\[E\: |\: \rho = S_0 \ar{\tau } S_1\ar{\tau } \ldots  \ar{\tau}S_i 
  \:[\ar{\tau }\ldots]
\]
  there always exists
\[S_1= !_{v^0_1}a\: |\: {\bar a}_{v^1_5}\: |\:Q(v^1_{2}, v^1_{3})\: 
  |\:a_{v_4}.\omega.
\]

\item  $v_4\not\in Ll(S_0)$,   $v_4\in Ll(S_1)$ and  $v_4\in
  Ll(S_{j+1})$ whenever there exists $k\geq (j+1)$ such that     
  ${\bar a}_{v^{k}_5}$  is a top-level parallel component of $S_{j+1}$.
\end{enumerate}
By definition of $Q(v^j_{2}, v^j_{3})$, there exist infinitely many
indexes $k$ such that ${\bar a}_{v^k_5}$ is a top-level parallel
component of $S_{j+1}$; it follows that $v_4$ can be live infinitely
often. But this is not possible if ${\mathcal C'}$ is a {\it
  strong-fair} computation: in fact, by definition, $v_4$ will lose
its liveness forever, i.e. $a_{v_4}$ will be performed.  In such a
case there will be $i\geq 2$ in ${\mathcal C'}$ such that
$S_{i}\ar{\omega}$.  \qed

Proposition \ref{risultati1} shows  the relation between
 {\it strong-fair must}- (respectively, {\it weak-fair must}-) testing semantics and {\it must}-testing.

\begin{prop}\label{risultati1} $\forall E \in {\mathcal P}^e.$ $\forall\rho \in {\mathcal O}^e$. 
\begin{enumerate}[\em1.]
\item  $Unl(E)\must Unl(\rho)$ implies $E \wfmust \rho$, but
not vice versa;

\item $Unl(E)\must Unl(\rho)$ implies $E \sfmust \rho$, but
not vice versa.
\end{enumerate}
\end{prop}

\proof\hfil
\begin{enumerate}[(1)]
\item For the positive part, suppose there is a {\it weak-fair} computation from $E\: |\:\rho$ 
\[ E\: |\:\rho=S_0\rid S_1\rid\dots\rid S_i\:[\rid\dots]\]
such that $\forall i\geq 0.\: S_i\not\ar{\omega}$. Then there exists the following maximal computation
\[Unl(E\: |\:\rho)=Unl(S_0)\rid Unl(S_1)\rid\dots\rid Unl(S_i)\:[\rid\dots]\]
where $\forall i\geq 0. \:Unl(S_i)\not\ar{\omega}$, i.e. $Unl(E)\!\notmust Unl(\rho)$.

We now prove the negative part. Let $E=(\nu b)({\bar b}_{v^0_1}\: |\:!_{v^0_2}b.{\bar b})\: |\:{\bar a}_{v_3}$
and $\rho = a_{v_4}.\omega$, we have $Unl(E)\notmust Unl(\rho)$. However, in 
every {\it weak-fair} computation from $E\: |\: \rho$
\[E\: |\:\rho=S_0\rid S_1\rid\dots\rid S_i\rid\dots\]
there must exist $j\geq 0$ such that 
$S_{j+1}=(\nu b)({\bar b}_{v^{j}_1} \: |\:!_{v^{j}_2}b.{\bar b})\: |\:\omega\ar{\omega}$ and
$\forall i\in[0..j].\: S_i= (\nu b)({\bar b}_{v^i_1}\: |\:!_{v^i_2}b.{\bar b})\: |\:{\bar a}_{v_3}\: |\:
a_{v_4}.\omega$.  It follows by the fact that $\forall i\in[0..j].\:  v_4\in Ll(S_i)$
and there must exist $k>i$ ($k=j+1$)  such that $v_4\not\in Ll(S_k)$. It is possible only in the case
$a_{v_4}.\omega$ synchronizes with ${\bar a}_{v_3}$ in $S_{k-1}=
(\nu b)({\bar b}_{v^k_1}\: |\:!_{v^k_2}b.{\bar b})\: |\:{\bar a}_{v_3}\: |\:
a_{v_4}.\omega$.

\item Immediate consequence of item (1) and Proposition \ref{risultatoclou}.
\qed
\end{enumerate}

\section{Fair-testing and `fair'-testing semantics}\label{omegafair}

In \cite{RV07} it is shown that {\it fair}-testing semantics on finite
state systems corresponds to some (strong) notion of
fairness. However, this result does not hold in general.  We will show
that {\it strong-fair must}-testing (and hence {\it weak-fair
  must}-testing) does not suffice to characterize {\it fair}-testing.

The reason behind the negative result relies on the fact that we can
construct a term for which there exist experiments being successful
under {\it fair}-testing and performing maximal unsuccessful
computations which are strong fair.

\begin{thm}\label{risultati2} $\forall E \in {\mathcal P}^e.$ $\forall\rho \in {\mathcal O}^e$.
\begin{enumerate}[\em 1.]
\item  $E\:\sfmust\: \rho\:$ implies $Unl(E) \fair Unl(\rho)$, but
not vice versa;

\item $E\:\wfmust \:\rho$ implies $Unl(E) \fair Unl(\rho)$, but
not vice versa.
\end{enumerate}
\end{thm}

\proof\hfil
\begin{enumerate}[(1)]
\item For the positive result, suppose, by contradiction, 
  there exists a maximal computation from  $Unl(E) \: |\:Unl(\rho)$
\[Unl(E) \: |\:Unl(\rho)=T_0\rid T_1\rid\dots\rid T_i\:[\rid\dots]\]
  and there exists $i\geq 0$ such that $T_i\not\Ar{\omega}$, i.e. for
  each $ T'$ such that $T_i\Ar{\varepsilon} T'$, we have
  $T'\not\ar{\omega}$. It follows that for every maximal computation
  from $T_i$ of the form
\[T_i=T'_0\rid T'_1\rid\dots\rid T'_j\:[\rid\dots]\]
  $T'_j\not\ar{\omega}$ for every $j$. Since $\omega$ cannot
  synchronize, it does not disappear once it is at the top level of a
  term. It implies that $\forall j\in[0..(i-1)].\:
  T_j\not\ar{\omega}$. Now, consider the computation
\[E\: |\:\rho=S_0\rid S_1\rid\dots\rid S_i\:[\rid\dots]\]
  where for every $ k\geq 0$ we have $T_k=Unl(S_k)$. Then there exists
  $i\geq 0$ such that $S_i\not\Ar{\omega}$, i.e.  for each $S'$ such
  that $S_i\Ar{\varepsilon} S'$, we have $S'\not\ar{\omega}$. It
  follows that for any maximal computation from $S_i$
\[S_i=S'_0\rid S'_1\rid\dots\rid S'_j\:[\rid\dots]\]
  $S'_j\not\ar{\omega}$ for every $j$. Hence for every {\it
  strong-fair} computation from $S_i$ 
  (which always exists, by Theorem \ref{strongweak})
\[S_i=S'_0\rid S'_1\rid\dots\rid S'_j\:[\rid\dots]\]
  $S'_j\not\ar{\omega}$ for every $j$. It follows that, given a 
  {\it strong-fair} computation from $S_i$
\[S_i=S''_0\rid S''_1\rid\dots\rid S''_j\:[\rid\dots]\]
  where $S''_j\not\ar{\omega}$ for every $j$, the following maximal
  computation
\[E\: |\:\rho=S_0\rid S_1\rid\dots\rid S_i=S'_0\rid S''_1\rid\dots\rid
  S''_j\:[\rid\dots]
\]  
  is {\it strong-fair} (by Lemma \ref{lemma1}), and $\forall
  k\in[0..(i-1)].\: S_k\not\ar{\omega}$, and $\forall j\geq 0.\:
  S''_j\not\ar{\omega}$. It follows that $E\notsfmust \rho$,
  contradicting the hypothesis.

  We now prove the negative part. As explained before, it suffices to
  consider $E= {\bar c}_{v^0_1}\: |\:!_{v^0_2}c.((\nu b)({\bar b}\:
  |\: b.{\bar c}\: |\:b.{\bar a}))$ and $\rho=
  a_{v_3}.\omega$. Clearly, $Unl(E) \:\fair\: Unl(\rho)$, but there
  exists the following maximal computation
\[\eqalign{
E\:|\:\rho={\bar c}_{v^0_1}\: |\:!_{v^0_2}c.((\nu b)({\bar b}\: |\:
  b.{\bar c}\: |\:b.{\bar a}))\:| \: a_{v_3}.\omega&\rid\cr
(\nu b)({\bar b}_{v^1_4}\: |\: b_{v^1_5}.{\bar c}_{v^1_6}\:
  |\:b_{v^1_7}.{\bar a}_{v^1_8})\: |\: !_{v^1_2}c.((\nu b)({\bar b}\:
  |\: b.{\bar c}\: |\:b.{\bar a}))\: |\:a_{v_3}.\omega&\rid\cr
(\nu b)(b_{v^1_7}.{\bar a}_{v^1_8})\: |\:{\bar c}_{v^1_6}\:
  |\:!_{v^1_2}c.((\nu b)({\bar b}\: |\: b.{\bar c}\: |\:b.{\bar a}))\:
  |\:a_{v_3}.\omega&\rid\cr
\ldots&\rid\cr
\prod_{i\in[1..k]}(\nu b)(b_{v^i_7}.{\bar a}_{v^i_8})\: |\:{\bar
  c}_{v^k_6}\: |\:!_{v^k_2}c.  ((\nu b)({\bar b}\: |\: b.{\bar c}\:
  |\:b.{\bar a}))\: |\:a_{v_3}.\omega&\rid\cr
\ldots&\rid\cr}
\]
where no term has $\omega$ enabled.  Notice that $\omega$ is always
prefixed in $a_{v_3}.\omega$ and $v_3$ is always disabled since every
occurrence of ${\bar a}_{v^i_8}$ is prefixed in a deadlock term $(\nu
b)(b_{v^i_7}.{\bar a}_{v^i_8})$.  Hence this computation is {\it
  strong-fair}.

\item The positive part is an immediate consequence of item (1) and Proposition \ref{risultatoclou}. As for the negative part, observe that the counterexample in the proof of item (1) is a counterexample here too, because the computation considered is also \emph{weak-fair}.\qed
\end{enumerate}

\noindent Previous result establishes that the notion of {\it weak}- and {\it
  strong-fair must}-testing differ from the notion of {\it
  fair}-testing in literature. A natural question is, then, which
notion is more suitable than the other in given situations.  As shown
by the counterexample in the proof of previous theorem, the difference
is with respect to computations that are fair but unsuccessful, and
they offer at every state the possibility of being successful. These
computations are considered acceptable by the notion of {\it
  fair}-testing, but not by our notion, and in our opinion, they
should not be.

\begin{exa}
We illustrate the difference with the well-known example of the dining philosophers. We can specify the system 
in our language in the following way. The system, \emph{DP}, is composed by three forks $\bar f_0,\bar f_1,\bar f_2$ and three 
philosophers $P_0,P _1,P_2$, in parallel: 
\begin{eqnarray*}
\emph{DP} &\stackrel{\rm def}{=}&   \bar f_0\: | \:  P_0\:  | \: \bar f_1\:  |\:  P _1\:  | \: \bar f_2\:   |\:  P_2
\end{eqnarray*}
Each philosopher replicates the following activity: first, he chooses  whether to start with the left fork (if available) or with the right fork (if available). For the choice we use the input-guarded choice construct, represented here by the operator $+$. It is well-known that this kind of choice can be expressed in the asynchronous $\pi$-calculus, and therefore also in the language that we consider here, by a translation that preserves must semantics \cite{NP00}. 
\[
P_i   \stackrel{\rm def}{=}   ! \: (L_i   +   R_{i})
\]
Under the left choice the philosopher takes the left fork, then chooses whether to take the right fork (if available) or to give up. In the first case, he takes the fork, eats, and then releases both forks. In the second case, he releases the left fork. This behavior can be represented as follows (where $\oplus$ denotes summation modulo $3$): 
\[
L_i    \stackrel{\rm def}{=}     f_i .  (f_{i\oplus 1} . \overline {\it eat} . (\bar f_i \: | \: \bar f_{i+1}) \:\: +  \:\: \tau. \bar f_i) 
\]
The behavior under the right choice is analogous:
\[
R_i    \stackrel{\rm def}{=}     f_{i\oplus 1} .  (f_{i} . \overline {\it eat}  . (\bar f_{i \oplus 1} \: | \: \bar f_{i}) \:\: +  \:\: \tau. \bar f_{i\oplus 1}) 
\]
Let us consider the observer which detects whether one philosopher succeeds to eat:
\[o  \stackrel{\rm def}{=}  {\it eat}\,.\, \omega\]
We can see that 
\[\emph{DP}\:\: \emph{fair} \:\: o \]
In fact, in every computation either a philosopher succeeds in taking both forks, and in that case he eats and the observer is satisfied, or there is always the possibility that one fork becomes available and can be taken by a philosopher who has already another fork. 
On the other hand, the computation in which each philosopher in turn takes the right fork, releases it, then take the second fork, releases it, then take the right fork \ldots etc. is strongly fair, and unsuccessful. Hence we have 
\[\emph{DP} \:\:\notsfmust \:\: o \]
\end{exa}

The answer given by our semantics is consistent with the view in Distributed Computing, where fairness and progress (a generalization of success - in this case, the fact that someone will eventually eat) are distinct concepts, and the Dining Cryptographers are considered an example of the fact that the first (fairness) does not imply the latter (progress).

The difference between {\it fair}-testing and both {\it weak}- and {\it strong-fair must}-testing 
relies on the fact that the former is based on properties of the transition tree  and the latter are based on the notion of fairness. 

We will prove in fact that no notion based only on the transition tree can characterize  {\it strong-fair must}- and {\it weak-fair must}-testing. To this purpose, let us  recall the definition of ({strong}) \emph{bisimulation}.
\begin{defi}({\it Bisimulation})\label{bis:definition}
A \emph{bisimulation} is a binary relation $\mathcal{R}$ satisfying the
following: $P \:\mathcal{R}\: Q$ implies that:
\begin{enumerate}[1.]
\item  $P\ar{\mu} P'$ then  $\exists Q':  Q \ar{\mu} Q' \wedge  P'\:\mathcal{R}\: Q'$;

\item  $Q\ar{\mu} Q'$ then  $\exists P':  P \ar{\mu} P' \wedge  P'\:\mathcal{R}\: Q'$.
\end{enumerate}
\emph{Bisimilarity} $\sim$ is the largest bisimulation $\mathcal{R}$ such that 
$P\:\mathcal{R}\: Q$.
\end{defi}
We recall that bisimilarity is a congruence. 

We  now prove that $\sfmust$ and $\wfmust$ cannot be characterized by a notion that, like {\it fair}-testing, relies on the transition tree only. 

\begin{thm}\label{impossibility} $\exists E, F\in{\mathcal P}^e.$ $Unl(E)\sim Unl(F)$ but $E\not\approx_{\!\mbox{\it sat}} F$, where $\mbox{\it sat}\in\{\!\!\wfmust, \sfmust\!\}$. 
\end{thm}
\proof Let  
\[E = (\nu c)({\bar c}_{w^0_0}\:| \:!_{w^0_1}c.({\bar c}\: |\: \bar a))\: \:| \:\:(\nu
c)({\bar c}_{w^0_2}\: |\:!_{w^0_3}c.{\bar c})
\] 
  and
\[F = !_{v^0_0}((\nu b) (\bar b\: |\:b\: |\: b.\bar a))\: |\: (\nu c)
(\bar c_{v^0_1} \:| \: !_{v^0_2}c.\bar c).
\]
$E$ and $F$ are neither  $\sfmust$ nor $\wfmust$ equivalent, since 
the observer $\rho=a_{v_3}.\omega$ distinguishes $E$ and $F$ w.r.t. both
$\sfmust$ and $\wfmust$. In fact, every {\it strong-fair} (respectively, {\it weak-fair}) computation 
from $E\:|\:\rho$ forces the synchronization between ${\bar c}_{w^0_0}$ and $!_{w^0_1}c.({\bar c}\: |\: \bar a)$, i.e. the transition 
$(\nu c)({\bar c}_{w^0_0}\:| \:!_{w^0_1}c.({\bar c}\: |\: \bar a)) \rid 
\bar a_{w^1_4}\: |\:(\nu c)({\bar c}_{w^1_0}\:| \:!_{w^1_1}c.({\bar c}\: |\: \bar a))$ and it also forces the execution 
of $\bar a_{w^1_4}$ (or equivalently of $\bar a_{w^i_4}$ for some $i\geq 1$ such that $(\nu c)
({\bar c}_{w^{i-1}_0}\:| \:!_{w^{i-1}_1}c.({\bar c}\: |\: \bar a)) \rid 
\bar a_{w^{i-1}_4}\: |\:(\nu c)({\bar c}_{w^{i-1}_0}\:| \:!_{w^{i-1}_1}c.({\bar c}\: |\: \bar a))$ occurred in the computation).

It follows that there exists a transition in which $a_{v_3}$ is
performed, implying that there exists a term which has $\omega$
enabled.

This is not the case of the following {\it strong-fair} (and {\it weak-fair}) computation from $F\:|\:\rho$:
\[\eqalign{
F\:|\: \rho = !_{v^0_0}((\nu b) (\bar b \:|\: b\: |\: b.\bar a))\: |\: (\nu c) (\bar c_{v^0_1} \:| \: !_{v^0_2}c.\bar c)\: |\: a_{v_3}.\omega&\Ar{\varepsilon}\cr
(\nu b)(b_{v^1_4}.{\bar a}_{v^1_5})\: |\: !_{v^1_0}((\nu b) (\bar b\: |\: b \:|\: b.\bar a)) \: |\: 
(\nu c) (\bar c_{v^1_1} \:| \: !_{v^1_2}c.\bar c) \: |\: a_{v_3}.\omega&\Ar{\varepsilon}\cr
(\nu b)(b_{v^1_4}.{\bar a}_{v^1_5})\: |\: (\nu b)(b_{v^2_4}.{\bar a}_{v^2_5})\: |\: 
!_{v^2_0}((\nu b) (\bar b \:|\: b \:|\: b.\bar a)) \: |\: (\nu c) (\bar c_{v^2_1} \:| \: !_{v^2_2}c.\bar c) \: |\: a_{v_3}.\omega&\Ar{\varepsilon}\cr
\ldots&\Ar{\varepsilon}\cr
\prod_{i\in[1..k]}(\nu b)(b_{v^k_4}.{\bar a}_{v^k_5})\: |\: 
!_{v^k_0}((\nu b) (\bar b\: |\: b |\: b.\bar a)) \: |\: (\nu c) (\bar c_{v^k_1} \:| \: !_{v^k_2}c.\bar c) \: |\: a_{v_3}.\omega&\Ar{\varepsilon}\cr
\ldots&\Ar{\varepsilon}
}\]
where there are no terms with $\omega$ enabled. Notice that $\omega$ is always
prefixed in $a_{v_3}.\omega$  and $a_{v_3}$ is
always disabled since  every occurrence of 
${\bar a}_{v^i_5}$ is prefixed in a deadlock term $(\nu b)(b_{v^i_4}.{\bar a}_{v^i_5})$. 
\par\medskip

However $Unl(E)\sim Unl(F)$, implying that $(Unl(E)\: |
\:Unl(\rho))\sim (Unl(F)\: | \:Unl(\rho))$, for any observer $\rho$. 
\qed

\section{Conclusion and future  work}\label{fur}

We have designed a labeled version of the $\pi$-calculus, we have defined 
weak and strong fairness, and we have 
introduced the natural  (weak and strong) fair versions of testing semantics. 
We have compared
the various notions and proved that neither weak nor strong fairness correspond to {\it fair}-testing, and we have investigated 
the reason of this failure. 

Our results are quite general, since they  also hold for 
CCS, for the asynchronous $\pi$-calculus \cite{Bou92} (it is easy to see that all proofs can be adapted immediately to these other calculi), to a  $\pi$-calculus with  
choice operator (as  explained in the introduction), and they do not depend on the labeling method (i.e. they hold 
for any labeling method for which  unicity, disappearance and  persistence 
 hold).

 As a future work, we plan to investigate on the existence of
alternative characterizations of the fairness notions,
allowing simple and finite representations of fair computations
such as the use of regular expressions as in \cite{CDV03,CDV04}. It
is also interesting to investigate the impact that these
different notions of fairness may have on the encodings from the
$\pi$-calculus into the asynchronous $\pi$-calculus \cite{CCP05}.

Another line of research that seems worth exploring is the the adaptation in our framework of the 
fairness notions of \cite{KdR83}. As we have mentioned in the introduction, it is possible to represent 
several forms of choice in the choiceless $\pi$-calculus using the parallel operator, and it would be 
interesting to see how the  fairness notions of \cite{KdR83} relative to the choice operator 
get translated in our formalism.

\section*{Acknowledgemnent}
We wish to thank the anonymous reviewers for their valuable comments and suggestion
which helped to improve the paper in a substantial way.

\appendix
\section{A labeled version of the \texorpdfstring{$\pi$}{pi}-calculus}\label{app:a}

This appendix section contains intermediate results and proofs of
the statements omitted in Section \ref{lablang}. Several
proofs follow the same lines as the corresponding results in
\cite{CS87}.

\begin{lem}\label{cat0}
Let $r_0, r_1,  s\in \{0,1\}^*.$ $r_0 \sqsubseteq s$ and $r_1 \sqsubseteq s$. 
Then either $r_0\sqsubseteq r_1$ or $r_1\sqsubseteq r_0$.
\end{lem}
\proof For $i\in\{0,1\}$, $r_i \sqsubseteq s$ implies $s=r_i\alpha_i$ for some 
$\alpha_i\in \{0,1\}^*$. Then $r_0\alpha_0=s =r_1\alpha_1$. Let $|r_i|$ the length of $r_i$. If $|r_0|\leq |r_1|$, 
then $r_0\sqsubseteq r_1$. Otherwise,  $r_1\sqsubseteq r_0$.
\qed

\begin{lem}\label{cat1} Let $\langle r_0, m_0\rangle, \langle r_1, m_1\rangle$, 
$\langle r_0', m_0'\rangle, \langle r_1', m_1'\rangle\in (\{0,1\}^*\times \N).$   
$r_0 \sqsubseteq r_0'$, $r_1 \sqsubseteq r_1'$ and $\{\langle r_0, m_0\rangle\}\Re \{\langle r_1, m_1\rangle\}$. Then 
$\{\langle r_0', m_0'\rangle\}\Re \{\langle r_1', m_1'\rangle\}$.
\end{lem}
\proof For $i\in\{0,1\}$, $r_i \sqsubseteq r_i'$ implies $r_i'=r_i\alpha_i$ for some 
$\alpha_i\in \{0,1\}^*$. By contradiction, suppose $r_0'\sqsubseteq r_1'$ (the other case is similar). Then 
$r_0\alpha_0\sqsubseteq r_1\alpha_1.$ Let $|r_i|$ the length of $r_i$. In the case $|r_0|\leq |r_1|$, then $r_0\sqsubseteq r_1$, contradicting 
$\{\langle r_0, m_0\rangle\}\Re \{\langle r_1, m_1\rangle\}$. In the case $|r_1|\leq |r_0|$, then $r_1\sqsubseteq r_0$, contradicting 
again $\{\langle r_0, m_0\rangle\}\Re \{\langle r_1, m_1\rangle\}$.
\qed

\begin{lem}\label{star} Let $E= L_{\langle r,m\rangle}(P)$, for some
$P\in {\mathcal P}$. $\forall \langle s,n\rangle \in lab(E)$. $r\sqsubseteq s$ and $m\leq n$.
\end{lem}
\proof
By induction on the structure of $P$.
\begin{enumerate}[--]
\item $E=0$: then $lab(0)=\emptyset$;

\item $E=L_{\langle r,m\rangle}(\mu.P)$: then
$lab(E)=\{\langle r,m\rangle\}\cup
lab(L_{\langle r,m+1\rangle}(P))$.

\item $E=L_{\langle r,m\rangle}(P_0\: |\: P_1)$: $lab(L_{\langle r,m\rangle}(P_0\: |\: P_1))
=lab(L_{\langle r0,m\rangle}(P_0))\cup lab(L_{\langle r1,m\rangle}(P_1))$. By
induction, $\forall \langle s_0,n_0\rangle \in lab(L_{\langle
r0,m\rangle}(P_0))$. $r\sqsubseteq r0\sqsubseteq s_0$ and $m\leq n_0$.
Analogously, $\forall \langle s_1,n_1\rangle \in lab(L_{\langle
r1,m\rangle}(P_1))$. $r\sqsubseteq r1\sqsubseteq s_1$ and $m\leq n_1$.

\item $E=!_{\langle r,m\rangle}P$: then $lab(E)=\{\langle
r,m\rangle\}$.

\item Case $E=L_{\langle r,m\rangle}((\nu x)P)$  can be proved similarly.\qed
\end{enumerate}

\begin{lem}\label{starstar} $\forall P\in {\mathcal P}.\: \forall \langle r, m\rangle\in (\{0,1\}^*\times \N)$. 
$\mathit{wf}(L_{\langle r,m\rangle}(P))$.
\end{lem}
\proof
By induction on the structure of $P$.
\begin{enumerate}[--]
\item $P=0, \mu.P',!P' $: these cases are trivial.

\item $P=P_0\: |\: P_1$:  then  $L_{\langle r,m\rangle}(P_0\: |\: P_1)=
L_{\langle r0,m\rangle}(P_0)\: |\: L_{\langle r1,m\rangle}(P_1)$ and by Lemma \ref{star} on 
$\topp(L_{\langle ri,m\rangle}(P_i))$ we have that $\forall \langle s_i,n_i\rangle \in 
\topp(L_{\langle ri,m\rangle}(P_i)).$  $ri\sqsubseteq s_i$ 
and $m\leq n_i$ ($i\in\{0,1\}$). Hence $\topp(L_{\langle r0,m\rangle}(P_0))\:\Re \:\topp(L_{\langle r1,m\rangle}(P_1))$.

\item $P=(\nu x)P'$: $L_{\langle r,m\rangle}(P)= (\nu x)L_{\langle r,m\rangle}(P')$, where 
$\mathit{wf}(L_{\langle r,m\rangle}(P'))$. Hence $\mathit{wf}(L_{\langle r,m\rangle}(P))$.\qed
\end{enumerate}

\begin{lem}\label{3-l5555}
Let $E\in {\mathcal P}^e$.   
\begin{enumerate}[\em1.]
\item For any distinct  $\langle r,m\rangle, \langle r',m'\rangle \in \topp(E)$.   
 $\{\langle r,m\rangle\}\Re \{\langle r',m'\rangle\}$;

\item $\forall \langle s,n\rangle \in lab(E)$.
$\exists  \langle r,m\rangle \in \topp(E).$ $r\sqsubseteq s$ and $m\leq n$.
\end{enumerate}
Let $E'\in {\mathcal P}^e_{\!{\it gr}}.$ $E\ar{\mu}{}E'$. Then: 
\begin{enumerate}[\em1.]
\item[\em3.] $\forall \langle r',m'\rangle\in \topp(E')$. $\exists
\langle r,m\rangle\in \topp(E).$ $r\sqsubseteq r'$ and $m\leq m'$;

\item[\em4.] $E'\in {\mathcal P}^e$.
\end{enumerate}
\end{lem}

\proof\hfil
\begin{enumerate}[(1)]
\item By induction on the structure of $E$.
\begin{enumerate}[--]
\item $E=0$: $\topp(0)=\emptyset$.

\item $E=L_{\langle s,n\rangle}(\mu.P)$: then $\topp(E)=\{\langle
s,n\rangle\}$.

\item $E=(E_0\: |\: E_1)$: since $\mathit{wf}(E_0 \:|
\:E_1)$ then $\topp(E_0)\Re \topp(E_1)$. Moreover, by induction hypothesis, 
$\forall \langle r_0,m_0\rangle, \langle r_0',m_0'\rangle \in \topp(E_0)$. 
$\{\langle r_0,m_0\rangle\}\Re \{\langle r_0',m_0'\rangle\}$ and, similarly, 
$\forall \langle r_1,m_1\rangle, \langle r_1',m_1'\rangle \in \topp(E_1)$. 
$\{\langle r_1,m_1\rangle\}\Re \{\langle r_1',m_1'\rangle\}$.

\item Case $E=(\nu x)E'$: it  can be proved similarly.

\item $E=!_{\langle s,n\rangle}P$:  then $\topp(E)=\{\langle
s,n\rangle\}$.
\end{enumerate}

\item By induction on the structure of $E$.
\begin{enumerate}[--]
\item $E=0$: $\topp(0)=\emptyset$ and $lab(0)=\emptyset$.
\par\smallskip
\item $E=L_{\langle s,n\rangle}(\mu.P)$: then $\topp(E)=\{\langle
s,n\rangle\}$. By Lemma \ref{star} on $L_{\langle s,n\rangle}(\mu.P)$, 
$\forall \langle s',n'\rangle \in lab(L_{\langle s,n\rangle}(\mu.P))$. 
$s\sqsubseteq s'$ and $n\leq n'$.

\item $E=(E_0\: |\: E_1)$: By induction, $\forall \langle
s_0,n_0\rangle\in lab(E_0). \:\exists\langle r_0,m_0\rangle\in
\topp(E_0) . $  $r_0\sqsubseteq s_0$ and $m_0\leq n_0$. Analogously
$\forall \langle s_1,n_1\rangle\in lab(E_1). \:\exists \langle
r_1,m_1\rangle\in \topp(E_1).$  $r_1\sqsubseteq s_1$ and $m_1\leq n_1$.  It follows that 
$\forall \langle s',n'\rangle\in lab(E_0\: |\: E_1)=lab(E_0)\cup lab(E_1).\:\exists\langle r',m'\rangle\in
\topp(E_0\: |\: E_1)= \topp(E_0)\cup \topp(E_1) .$  $r'\sqsubseteq s'$ and $m'\leq n'$. 

\item Case $E=(\nu x)E'$: it  can be proved similarly.

\item $E=!_{\langle s,n\rangle}P$:  then $\topp(E)=\{\langle
s,n\rangle\}=lab(E)$.
\end{enumerate}

\item By induction on the depth of $E\ar{\mu}{}E'$.
\begin{enumerate}[--]
\item {\it Rule Input/Output}: $E=L_{\langle
s,n\rangle}(\mu.P)\ar{\mu}{}E'=L_{\langle s,n+1\rangle}(P')$
(either $P' = P$ or $P' = P\{z/y\}$). Then $\topp(L_{\langle
s,n\rangle}(\mu.P))=\{\langle s,n\rangle\}$. By Lemma \ref{star} 
on $L_{\langle s,n+1\rangle}(P')$, we have that $\forall \langle r',m'\rangle \in 
\topp(L_{\langle s,n+1\rangle}(P'))\subseteq lab(L_{\langle s,n+1\rangle}(P'))$.
$s\sqsubseteq r'$ and $n+1\leq m'$. It follows that $\forall\langle
r',m'\rangle\in \topp(L_{\langle s,n+1\rangle}(P'))$.  
$s\sqsubseteq r'$ and $n<  m'$. 

\item {\it Rule Par}: $E=(E_0 \:| \:E_1) \ar{\mu} (E_0'\:| \:E_1)$,
where $bn(\mu)\cap fn(E_1) = \emptyset$ and $E_0 \ar{\mu} E_0'$. Since $\mathit{wf}(E_0 \:|
\:E_1)$, then $\topp(E_0)\:\Re\: \topp(E_1)$. 
By induction, $E_0 \ar{\mu} E_0'$ implies that $\forall \langle
r_0',m_0'\rangle\in \topp(E_0')$. $\exists\langle r_0,m_0\rangle\in
\topp(E_0).$ $r_0\sqsubseteq r_0'$ and $m_0\leq m_0'$. 
Since  $\topp(E_0\:| \:E_1)=\topp(E_0)\cup \topp(E_1)$ and $\topp(E_0'\:| \:E_1)=\topp(E_0')\cup \topp(E_1)$, then 
$\forall \langle r',m'\rangle\in \topp(E'_0\:| \:E_1).$ either $\exists \langle r_0,m_0\rangle\in \topp(E_0).$ 
$r_0\sqsubseteq r'$ and $m_0\leq m'$ (in the case $\langle r',m'\rangle\in \topp(E_0')$) or 
$\exists \langle r_1,m_1\rangle\in \topp(E_1).$ $r_1= r'$ and $m_1= m'$ (in the case $\langle r',m'\rangle\in \topp(E_1)$).

\item {\it Rule Com}: $E=(E_0 \:| \:E_1) \ar{\tau} (E_0'\:| \:E_1')$,
where $E_0 \ar{xy} E_0'$ and $E_1 \ar{{\bar x}y}{} E_1'$. 
By induction hypothesis, $\forall \langle r_0',m_0'\rangle\in \topp(E_0')$. $\exists
\langle r_0,m_0\rangle\in \topp(E_0).$ $r_0\sqsubseteq r_0'$ and $m_0\leq m_0'$. Analogously, 
$\forall \langle r_1',m_1'\rangle\in \topp(E_1')$. $\exists
\langle r_1,m_1\rangle\in \topp(E_1).$ $r_1\sqsubseteq r_1'$ and $m_1\leq m_1'$. 
Since  $\topp(E_0\:| \:E_1)=\topp(E_0)\cup \topp(E_1)$ and $\topp(E_0'\:| \:E_1')=\topp(E_0')\cup \topp(E_1')$, 
then $\forall \langle r',m'\rangle\in \topp(E_0'\:| \:E_1')$ either $\exists \langle r_0,m_0\rangle\in \topp(E_0).$ 
$r_0\sqsubseteq r'$ and $m_0\leq m'$ (in the case $\langle r',m'\rangle\in \topp(E_0')$) or 
$\exists \langle r_1,m_1\rangle\in \topp(E_1).$ 
$r_1\sqsubseteq r'$ and $m_1\leq m'$ (in the case $\langle r',m'\rangle\in \topp(E_1')$).

\item {\it Rule Open/Res/Close}: These cases can be proved similarly.

\item {\it Rule Rep}: $!_{\langle s,n\rangle}P \ar{\mu}{}
L_{\langle s0,n+1\rangle}(P) \:| \: !_{\langle
s1,n+1\rangle}P$. Then we have $\topp(!_{\langle s,n\rangle}P)\!=\!\{\langle s,n\rangle\}$ and
$\topp(L_{\langle s0,n+1\rangle}(P)\! | !_{\langle s1,n+1\rangle}P)$ 
$=\topp(L_{\langle s0,n+1\rangle}(P))\cup\{\langle s1,n+1\rangle\}$. By Lemma \ref{star} 
on $L_{\langle s0,n+1\rangle}(P)$, we have that $\forall \langle r',m'\rangle \in 
\topp(L_{\langle s0,n+1\rangle}(P))\subseteq lab(L_{\langle s0,n+1\rangle}(P))$.
$s0\sqsubseteq r'$ and $n+1\leq m'$. It follows that $\langle s,n\rangle$ is such that 
$s\sqsubseteq s1$ and $n< n+1$, as well as $s\sqsubseteq s0\sqsubseteq r'$ and $n< n+1\leq m'$ for any $\langle
r',m'\rangle\in \topp(L_{\langle s0,n+1\rangle}(P))$.  
\end{enumerate}

\item We prove that $\mathit{wf}(E')$ holds, by induction on the depth of $E\ar{\mu}{}E'$.
\begin{enumerate}[--]
\item {\it Rule Input/Output}: $E=L_{\langle
s,n\rangle}(\mu.P)\ar{\mu}{}E'=L_{\langle s,n+1\rangle}(P')$
(either $P' = P$ or $P' = P\{z/y\}$). By Lemma \ref{starstar},
$\mathit{wf}(L_{\langle s,n+1\rangle}(P'))$.

\item {\it Rule Par}: $E=(E_0 \:| \:E_1) \ar{\mu} (E_0'\:| \:E_1)$,
where $bn(\mu)\cap fn(E_1) = \emptyset$ and $E_0 \ar{\mu} E_0'$. Since $\mathit{wf}(E_0 \:|
\:E_1)$, then $\topp(E_0)\:\Re\: \topp(E_1)$. By induction, $E_0 \ar{\mu} E_0'$
implies that $\mathit{wf}(E_0')$. By item (3) and Lemma \ref{cat1},  $\topp(E_0)\:\Re\: \topp(E_1)$ 
implies $\topp(E_0')\:\Re\:\topp(E_1)$. Hence $\mathit{wf}(E_0' \:| \:E_1)$.

\item {\it Rule Com}: $E=(E_0 \:| \:E_1) \ar{\tau} (E_0'\:| \:E'_1)$,
where $E_0 \ar{xy} E_0'$ and $E_1 \ar{{\bar x}y}{} E_1'$. 
By induction hypothesis,  $E_0 \ar{xy} E_0'$
implies that $\mathit{wf}(E_0')$; analogously, $E_1 \ar{{\bar x}y}{} E_1'$ implies that 
$\mathit{wf}(E_1')$. By item (3) and Lemma \ref{cat1},  $\topp(E_0)\:\Re\: \topp(E_1)$ 
implies $\topp(E_0')\:\Re\:\topp(E_1')$. Hence $\mathit{wf}(E_0' \:| \:E_1')$.

\item {\it Rule Open/Res/Close}: These cases can be proved similarly.

\item {\it Rule Rep}: It suffices to recall that
$\topp(L_{\langle s0,n+1\rangle}(P')\! | !_{\langle s1,n+1\rangle}P')$
 $=\{\langle s1,n+1\rangle\}\cup \topp(L_{\langle s0,n+1\rangle}(P'))$ and to apply Lemma \ref{star} on  
 $L_{\langle s0,n+1\rangle}(P')$.\qed
 \end{enumerate}
\end{enumerate}

\begin{lem}\label{add1} Let $E_0, E_1\in {\mathcal P}^e$. 
$\topp(E_0)\:\Re\: \topp(E_1)$ implies $lab(E_0)\:\Re\: lab(E_1).$
\end{lem}
\proof By item (2) of Lemma \ref{3-l5555}, $\forall i\in\{0,1\}$. 
$\forall \langle s_i,n_i\rangle \in lab(E_i)$.
$\exists  \langle r_i,m_i\rangle \in \topp(E_i) . $ $r_i\sqsubseteq s_i$ and $m_i\leq n_i$. 
$\topp(E_0)\:\Re\: \topp(E_1)$ and Lemma \ref{cat1} imply  $\forall \langle s_0,n_0\rangle \in lab(E_0)$. 
$\forall \langle s_1,n_1\rangle \in lab(E_1)$. $\{\langle s_0,n_0\rangle\}\Re \{\langle s_1,n_1\rangle\}$, i.e. 
$lab(E_0)\:\Re\: lab(E_1).$
\qed

\begin{lem}\label{add3} Let $E, E'\in {\mathcal P}^e.$ $E\ar{\mu}{}E'$. 
Let $\langle s,n\rangle\in lab(E)$ and $\langle s,n\rangle\not\in lab(E')$. Then  
$\langle s,n\rangle\in \topp(E)$.
\end{lem}
\proof
By induction on the depth of $E\ar{\mu}{}E'$.
\begin{enumerate}[--]
\item {\it Rule Input/Output}: $E=L_{\langle
s',n'\rangle}(\mu.P)\ar{\mu}{}E'=L_{\langle s',n'+1\rangle}(P')$
(where either $P' = P$ or $P' = P\{z/y\}$). Then $\topp(E)=\{\langle s',n'\rangle\}$
and $lab(E) = lab(E') \cup \{\langle s',n'\rangle\}$. It follows that $s = s', n=n'$, 
and therefore $\langle s,n\rangle\in \topp(E)$.

\item {\it Rule Par}: $E=(E_0 \:| \:E_1) \ar{\mu} (E_0'\:| \:E_1)$ and $E_0 \ar{\mu} E_0'$. 
Since $lab(E_0 \:|\:E_1)=lab(E_0)\:\cup\: lab(E_1)$ and 
$lab(E_0' \:|\:E_1)=lab(E_0')\:\cup\: lab(E_1)$, we have $\langle s,n\rangle\not\in lab(E_0')$, 
$\langle s,n\rangle\not\in lab(E_1)$, and therefore $\langle s,n\rangle\in lab(E_0)$. 
By induction, 
$\langle s,n\rangle\in \topp(E_0)$ and therefore $\langle s,n\rangle\in \topp(E_0)\:\cup\: \topp(E_1)= \topp(E_0 \:|\:E_1)= \topp(E)$.

\item {\it Rule Com}: $E=(E_0 \:| \:E_1) \ar{\tau} (E_0'\:| \:E_1')$,
where $E_0 \ar{xy} E_0'$ and $E_1 \ar{{\bar x}y}{} E_1'$. 
Since  $lab(E_0 \:|\:E_1)=lab(E_0)\:\cup\: lab(E_1)$, we have that either
$\langle s,n\rangle\in lab(E_0)$ or $\langle s,n\rangle\in lab(E_1)$. Let us consider the first case (the other one is analogous). 
Since $lab(E_0' \:|\:E_1')=lab(E_0')\:\cup\: lab(E_1')$, we have that $\langle s,n\rangle\not\in lab(E_0')$. 
The rest is the same as in the case of {\it Par}.

\item {\it Rules Open/Res}: Immediate, by induction.

\item {\it Rule Close}: Similar to the case of {\it Com}. 
\item {\it Rule Rep}: Trivial, since $E = !_{\langle s',n'\rangle}P$ and $lab(!_{\langle s',n'\rangle}P) = \{\langle s',n'\rangle\}
 =\topp(!_{\langle s',n'\rangle}P)$.\qed
\end{enumerate}

\begin{lem}\label{add2} Let $E, E'\in {\mathcal P}^e.$ $E\ar{\mu}{}E'$ and 
$\langle r,m\rangle\in \topp(E)$. $\langle r,m\rangle\in lab(E')$ implies 
$\langle r,m\rangle\in \topp(E')$.
\end{lem}
\proof Let $\langle r,m\rangle\in \topp(E)\cap lab(E')$ and suppose, by contradiction, 
that $\langle r,m\rangle\not\in \topp(E')$. By  item (2) of Lemma \ref{3-l5555}, $\exists  \langle r',m'\rangle \in \topp(E') . $ 
$r'\sqsubseteq r$ and $m'\leq m$. By item (3) of Lemma \ref{3-l5555}, $\exists  \langle r'',m''\rangle \in \topp(E) . $ 
$r''\sqsubseteq r'\sqsubseteq r$ and $m''\leq m'\leq m$. It follows that $\exists \langle r,m\rangle, \langle  r'',m''\rangle\in \topp(E)$ 
such that $r''\sqsubseteq r$ and $m''\leq m$. Note also that the pairs 
$  \langle r,m\rangle, \langle r',m'\rangle$  must be different and therefore also 
$ \langle r,m\rangle, \langle  r'', m''\rangle$ are different. 
Thus we get a contradiction with item (1) of Lemma \ref{3-l5555}.
\qed

\section{Must- and fair-testing semantics}\label{app:b}

This appendix section contains intermediate results and proofs of
the statements omitted in Section \ref{swf}.

\begin{prop}\label{mustimpfair}
Let $P \in {\mathcal P}$ and $o \in {\mathcal O}$. $P\must o$ implies $P \fair o$.
\end{prop}
\proof By contradiction, suppose $P\:\notfair\: o$, i.e. there is 
a maximal computation from $P\: |\:o$
\[ P\: |\:o=T_0\rid T_1\rid\dots\rid T_i\:[\rid\dots]\]
such that $T_i\not\Ar{\omega}$ for some $i\geq 0$, i.e. for every $T'.\:
T_i\Ar{\varepsilon}T'$ it holds that $T'\not\ar{\omega}$. It follows that
$T_i\not\ar{\omega}$, $\forall j\in[0..(i-1)].  T_j\not\ar{\omega}$ and $\forall h\geq i. \:  T_h\not
\ar{\omega}$, by hypothesis on $T_i$. In fact, since $\omega$ can not synchronize, it does not disappear
once it is at the top level of a term.  It follows that the above computation 
is such that  $\forall j\geq 0. \: T_j\not\ar{\omega}$, i.e. $P\notmust o$.
\qed

\begin{prop}\label{fairnotimpmust}
$\exists P \in {\mathcal P}.$ $\exists o \in {\mathcal O}.$ $P\fair o$ and $P \notmust o$.
\end{prop}
\proof Consider $P=(\nu b)({\bar b}\: | \:!b.{\bar b})\: |\:{\bar a}$ and $o=a.\omega$.
Since  $(\nu b)({\bar b}\: | \:!b.{\bar b})\rid(\nu b)({\bar b}\: | \:!b.{\bar b})\rid\dots$,  
there is an unsuccessful maximal computation from $P\:|\:o$, i.e. $P\notmust o$. However, $P\fair o$, since every maximal 
computation from $P\: |\: o$
\[ P\: |\:o=T_0\rid T_1\rid\dots\rid T_i\rid\dots \]
is such that either $\forall i\geq 0. \: T_i= (\nu b)({\bar b}\: |
\:!b.{\bar b})\: |\:{\bar a}\: |\:a.\omega$ or $\exists j\geq 1.$
$T_j= (\nu b)({\bar b}\: | \:!b.{\bar b})\: |\:\omega\ar{\omega}$ and
$\forall i\in[0..(j-1)]. \: T_i= (\nu b)({\bar b}\: | \:!b.{\bar b})\:
|\:{\bar a}\: |\:a.\omega$ and $T_i\Ar{\varepsilon}T_j$.  \qed

\section{Weak-fair must, strong-fair must and fair-testing semantics}\label{app:c}

\begin{lem}\label{lemma0} $\forall S\in {\mathcal E}^e$.
\begin{enumerate}[\em1.]
\item $Ll(S)$ is a finite set;

\item $S\not\ar{\tau}$ implies $Ll(S)=\emptyset$;

\item $v\in Ll(S)$ implies $\exists S'\in {\mathcal E}^e.$ $S\ar{\mu}S'$ and 
$\forall S''.$ $S'\Ar{\varepsilon} S''$. $v\not\in Ll(S'')$;

\item $\exists S'\in {\mathcal E}^e.$ $S\Ar{\varepsilon}S'$, $Ll(S)\cap Ll(S')=\emptyset$ and
 $\forall S''.$ $S'\Ar{\varepsilon} S''$. $Ll(S)\cap Ll(S'')=\emptyset$.
\end{enumerate}
\end{lem}

\proof We recall that $\forall S\in{\mathcal E}^e.\: Ll(S)\subseteq \topp(S)\subseteq lab(S)$.
Items (1) and  (2) are trivial. Consider item (3). $S'$ is the term obtained from $S$ by
performing the action labeled by $v$: by Theorem \ref{a} , $v\not\in lab(S')$ and for every
$ S''$ such that $S'\Ar{\varepsilon} S''$,  $v\not\in lab(S'')$ holds. Hence $v\not\in Ll(S')$
and for every $ S''$ such that $S'\Ar{\varepsilon} S''$,  $v\not\in Ll(S'')$ holds.

To prove item (4) it suffices to apply the previous item, where
$\mu=\tau$. The term $S'$ is obtained from $S$  by performing any
$v\in Ll(S)$ and such that for every $ v\in Ll(S)$ and every $S''$ such that 
$S'\Ar{\varepsilon} S''$ either $v\not\in lab(S')$
(following that $v\not\in lab(S'')$) or $v\not\in Ll(S'')$ and $v\in lab(S')$. In both cases, $Ll(S)\cap Ll(S')=\emptyset$
and  $Ll(S)\cap Ll(S'')=\emptyset$. Since $Ll(S)$ is finite, such $S'$ exists.
\qed

\begin{lem}\label{lemma1}
Let $S \in {\mathcal E}^e$ and $S=S_0\rid S_1\rid\dots\rid S_i\:[\rid\dots]$ be a {\it strong-fair} computation from $S$.
If $\exists S'_0, S'_1,S'_2,\dots, S'_n\in {\mathcal E}^e$ such that
\[ S'=S'_0\rid S'_1\rid\dots\rid S'_n=S, \]
then 
\[ S'\rid S'_1\rid\dots\rid S'_n\rid S_1\rid\dots\rid S_i\:[\rid\dots]\]
is a {\it strong-fair} computation from $S'$.
\end{lem}

\proof Consider  ${\mathcal C}= S'\rid S'_1\rid\dots\rid S'_n\rid S'_{n+1}\rid\dots\rid S'_{n+i}\:[\rid\dots]$,
where $\forall j \geq 0. \: S'_{n+j}::=S_j$. Obviously ${\mathcal C}$ is a maximal computation  from $S'$. 
To prove that ${\mathcal C}$ is also {\it strong-fair},
it suffices to prove that  $\forall v\in
(\{0,1\}^*\times\N) .$ $\exists h\geq 0 .$  $\forall k \geq h. \:v\not\in Ll(S'_k)$. Since
$S'_n\rid S'_{n+1}\rid\dots\rid S'_{n+i}\:[\rid\dots]$ is a {\it strong-fair} computation from $S'_n$, then
$\forall v\in (\{0,1\}^*\times\N).$ $\exists h\geq n.$ $\forall k \geq h.\:v\not\in Ll(S'_k)$.  
Since $n\geq 0$, $\forall v\in (\{0,1\}^*\times\N).$ $\exists h\geq 0.$  
$\forall k \geq h.\: v\not\in Ll(S'_k)$. I.e.,   
${\mathcal C}$ is a {\it strong-fair} computation from $S'$.
\qed


\begin{thebibliography}{}



\bibitem{AMST97} Agha, G., Mason, I. A., Smith, S. \& Talcott, C. L.,  
{\em A Foundation for Actor Computation}, 
JFP, {\bf 7}(1) (1997), 1-72.




\bibitem{BD95} Boreale, M. \&  De Nicola, R., {\em Testing Equivalence for Mobile Processes}, 
Information and Computation, {\bf 120} (1995), 279-303.

\bibitem{Bou92} Boudol, G., \emph{Asynchrony and the $\pi $-calculus'},
Technical Report 1702, INRIA, Sophia-Antipolis (1992).



\bibitem{Bri84} Brinksma, E., {\em Cache Consistency by Design}, 
In \emph{Protocols Specification,
Testing and Verification} (VIII), Aggarwal \& Sabnani (1988), 63-74.

\bibitem{Bri88} Brinksma, E., {\em A theory for the Derivation of Tests}, 
In ``Protocols Specification,
Testing and Verification'' (XIV), Chapman \& Hall (1995), 53-67.






\bibitem{BRV96}  Brinksma. E., Rensink, A. \& Vogler, W., 
\emph{Applications of Fair Testing}, In \emph{Protocols Specification,
Testing and Verification} (XVI), Chapman \& Hall (1996), 145-160.

\bibitem{CCP05} Cacciagrano, D.,  Corradini, F. \&  Palamidessi, C.,  
{\em Separation of Synchronous and Asynchronous Communication Via
Testing}, Theoretical  Computer Science, {\bf 386}(3) (2007), 218-235. 
 


\bibitem{CDV03} Corradini, F., Di Berardini, M.R. \&  Vogler, W., 
{\em Fairness of Actions in System Computations}, Acta Informatica, {\bf 43}(2) (2006), 73-130.

 
\bibitem{CDV04} Corradini, F., Di Berardini, M.R. \&  Vogler, W., 
{\em Fairness of Components in System Computations}, 
Theoretical  Computer Science, {\bf 356}(3) (2006), 291-324.



\bibitem{CS84} Costa, G. \& Stirling, C., 
{\em A Fair Calculus of Communicating Systems}, Acta Informatica,
{\bf 21} (1984), 417-441.

\bibitem{CS87} Costa, G. \& Stirling, C., {\em Weak and Strong Fairness in CCS}, 
Information and Computation, {\bf 73} (1987), 207-244.

\bibitem{DH84} De Nicola, R. \& Hennessy, M., 
{\em Testing Equivalences for Processes}, Theoretical  Computer Science, {\bf 34} (1984), 83-133.


\bibitem{FG98} Fournet, C. \& Gonthier, G., 
{\em A Hierarchy of Equivalences for Asynchronous Calculi}, Proc. of
ICALP'98 (1998), 844-855.

\bibitem{FG00} Fournet, C. \& Gonthier, G., 
{\em The Join Calculus: A Language for Distributed Mobile Programming}, Proc. of
APPSEM 2000, LNCS, {\bf 2395} (2000), 268-332.




\bibitem{Fra86} Francez, N., \emph{Fairness}, Springer-Verlag (1986).





\bibitem{Hen85} Hennessy, M., 
{\em Acceptance trees}, JACM, {\bf 32}(4) (1985), 896-928.



\bibitem{Hen87} Hennessy, M., 
{\em An Algebraic Theory of Fair Asynchronous Communicating
Processes}, Theoretical Computer Science, {\bf 49} (1987), 121-143.






\bibitem{KdR83} Kuiper, R. \& de Roever, W. P., {\em Fairness assumptions for CSP in a temporal logic
framework}, Proc. of IFIP Working Conference on
Formal Description of Programming Concepts (1983), 159-167. 


\bibitem{HT91} Honda, K. \&  Tokoro, M., {\em An Object calculus for Asynchronous Communication}, 
Proc. of ECOOP '91, LNCS,  {\bf 512} (1991), 133-147.

\bibitem{HY94} Honda, K. \&  Yoshida, N., 
{\em Replication in Concurrent Combinators}, Proc. of TACS '94, LNCS, {\bf 789} (1994).  

\bibitem{Koo85} Koomen, C., 
{\em Albegraic Specification and Verification of Communications
protocols}, Science of Computer Programming, {\bf 5} (1985), 1-36.

\bibitem{LM87} Larsen, K. G. \&  Milner, R., {\em Verifying a Protocol 
using Relativized Bisimulation}, LNCS, {\bf 267} (1987), 126-135.

\bibitem{LPS81} Lehmann, D., Pnueli, A. \& Stavi, J., 
{\em Impartiality, justice and Fairness:the Ethics of Concurrent
Termination}, Proc. of 8th Int. Colloq. Aut. Lang. Prog., LNCS,
{\bf 115} (1981), 264-277.

\bibitem{Mil89} Milner, R., \emph{Communication and Concurrency}, Prentice-Hall International (1989).

\bibitem{MPW92} Milner, R., Parrow, J. \& Walker, D.,  {\em A Calculus of Mobile
Processes}, Part I and II, Information and Computation, {\bf 100} (1992), 1-78.


\bibitem{NC95} Natarajan, V. \&  Cleaveland, R., {\em Divergence and Fair
Testing}, Proc. of ICALP '95, LNCS, {\bf 944} (1995), 648-659.

\bibitem{Nes00}
Nestmann, U. 
\newblock What is a `good' encoding of guarded choice? 
\newblock {\em Journal of Information and Computation}, 156:287--319, 2000.

\bibitem{NP00}
Nestmann, U. and Pierce, B.~C. 
\newblock Decoding choice encodings.
\newblock {\em Journal of Information and Computation}, 163:1--59, 2000.

\bibitem{NR99} N\'u$\tilde{\mbox{\rm n}}$ez, M. \& Rup\'erez, D., 
{\em Fair testing through probabilistic testing}, Protocol Specification,
Testing, and Verification, {\bf 19} (1999), 135-150.


\bibitem{Pal03}  Palamidessi, C.\enspace
{\it Comparing the Expressive Power of the Synchronous and
Asynchronous $\pi$-calculus}, Mathematical Structures in Computer
Science, {\bf 13}(5), pp. 685-719, 2003. 


\bibitem{Par80} Park, D. M. R., {\em Concurrency and Automata on Infinite Sequences},
 LNCS, {\bf 104} (1980).

\bibitem{Pnu83} Pnueli, A., {\em On the Extremely Fair Treatment 
of Probabilistic Algorithms}, Proc. of ACM Symph. Theory of Comp. (1983), 278-290.

\bibitem{PT00} Pierce, B. C. \& Turner, D. N., {\em Pict: 
A Programming Language Based on the Pi-Calculus}, in  
 \emph{Proof, Language and Interaction: Essays in Honour of Robin Milner}, MIT Press (2000).





\bibitem{QS83} Queille, J.P. \& Sifakis, J., {\em Fairness and Related 
Properties in Transition Systems-A
Temporal Logic to Deal with Fairness}, Acta Informatica, {\bf 19} (1983), 195-210.

\bibitem{RV07}  Rensink, A. \& Vogler, W., {\em Fair
Testing}, Information and Computation, {\bf 205} (2007), 125-198.
 A short version of this paper appeared in the Proc. of
CONCURÕ95, LNCS, 962 (1995), 313-327.


\bibitem{San98} Sangiorgi, D., {\em On the Bisimulation Proof Method}, 
JMSCS, {\bf 8} (1998), 447–479.


\bibitem{SW01} Sangiorgi, D. \& Walker, D., \emph{The Pi-calculus: a 
Theory of Mobile Processes}, Cambridge University Press (2001).


\bibitem{Vog92} Vogler, W., {\em Modular Construction and Partial Order Semantics of Petri Nets}, LNCS, {\bf 625} (1992). 
\end{thebibliography}
\end{document}